# Luminosity Evolution of the Hot Gas in Normal Galaxies from the Near Universe to z=0.5

Dong-Woo Kim & Giuseppina Fabbiano

Center for Astrophysics | Harvard and Smithsonian
60 Garden Street, Cambridge, MA 02138, USA

**Abstract**

We explore the evolution of the ~$10^7$ K hot gas in normal galaxies out to redshift = 0.5 (lookback time = 5 Gyr), using X-ray luminosity functions (XLF) built from a sample of 575 normal galaxies with z < 0.6 detected in five high galactic latitude Chandra wide-field surveys. After estimating the emission due to the hot gas component (reducing the sample to ~400 galaxies), we compared the XLF in three redshift bins (z = 0.1, 0.3, and 0.5), finding increases in the number of galaxies per unit co-moving volume from z = 0.1 to 0.3 and then from z = 0.3 to 0.5. These XLF changes suggest a significant (~5σ) X-ray luminosity evolution of the hot gas, with $L_{X,GAS}$ decreasing by a factor of 6-10 in the last 5 Gyr (from z = 0.5 to 0.1). The relative abundance of $L_{X,GAS}$~$10^{41}$ erg $s^{-1}$ galaxies at higher z, suggests that high z, moderate $L_{X,GAS}$ galaxies may be the optimal target to solve the missing baryon problem. In early-type galaxies, this observational trend is qualitatively consistent with (but larger than) the expected time-dependent mass-loss rate in cooling flow models without AGN feedback. In late-type galaxies, the observational trend is also qualitatively consistent with (but larger than) the effect of the z-dependent SFR.

## 1. INTRODUCTION

The formation and evolution of galaxies is reflected in the physical and chemical properties of their hot, X-ray-emitting, $\sim10^7$ K gaseous component. From X-ray observations of nearby galaxies (see review, Fabbiano 2020), we have learnt that intense star formation episodes may cause hot, metal enriched and often outflowing galaxy wide halos. Large hot halos are also gravitationally retained by the massive dark halos of early-type, non-rotating cluster galaxies, and can be affected by environmental effects such as infall, stripping, and mergers. Feedback from a nuclear active supermassive black hole may heat the interstellar medium, and there is growing evidence that small radio jets embedded in galaxy disks may also cause the formation of hot halos (see review, Fabbiano & Elvis 2022).

The hot $\sim10^7$ K gas may provide the working surface necessary for AGN feedback by storing and smoothing episodic energy input; it also may shield against the accretion of fresh cool gas, thus impeding star formation (Nipoti & Binney 2007; Gabor and Dave 2015). The observational measurements of the hot gas in local galaxies and the predictions from the simulations have been compared with some success, and have posed constraints on feedback recipes (e.g., Choi et al. 2014, Ciotti et al. 2017; Wang et al. 2023). These works have been focused on the nearby universe because of the observational limitations of the X-ray studies of galaxies. However, the history of the hot gas properties as a function of redshift may provide even stronger constraints on the interplay of galaxy formation and feedback.

Studying the E-CDF-S (extended Chandra deep field south), Lehmer et al. (2007) found that the X-ray to optical luminosity ratio, $L_X$ (0.5-2 keV) / $L_B$, does not change with z for large galaxies ($L_B > 10^{10}\ L_{B\odot}$), while this ratio increases with z for small galaxies ($L_B < 10^{10}\ L_{B\odot}$) at z=0-0.5. Analyzing the CDF data (CDF-N, CDF-S, and E-CDF-S), Danielson et al. (2012) found a mild evolution in $L_X$ (0.5-2 keV) / $L_B \sim (1+z)^{1.2}$. Jones et al. (2014) stacked the K-band limited, higher z (0.5-2) galaxies in the COSMOS field and found $L_X$ (0.5-7 keV) / $M_{STAR}$ increasing by an order of magnitude from z=0.5 to z=2 for all types of galaxies and more AGN contributions at higher z. Paggi et al. (2016) also stacked early-type galaxies (z < 1.5) in the COSMOS field and found a significant hidden AGN contribution at higher z. These previous studies provided mixed results on the luminosity evolution, but they included all the X-ray emission components in a wider energy band instead of focusing on the hot gas.

In quiescent early-type galaxies with no star formation and no AGN, the luminosity of the hot gas $L_{X,GAS}$ primarily depends on the accumulated mass loss from the evolved stars, which monotonically decreases with time (e.g., Pellegrini 2012). In normal (non-AGN) late-type galaxies, $L_{X,GAS}$ depends on the star formation rate (Mineo et al. 2012). In both early- and late-type galaxies, we would expect $L_{X,GAS}$ monotonically decreasing with age, i.e., higher $L_{X,GAS}$ at higher z.

However, stellar and AGN feedback may change this simple scenario. It is known that the X-ray luminosities of both low-mass (LMXBs) and high-mass X-ray binaries (HMXBs), increase with increasing redshifts (Fragos et al. 2013; Lehmer et al. 2016), consistent with the star formation rate (SFR) peaking at redshift z = 2-3 (Madau & Dickinson 2014). With increasing redshift (from z=0.1 to z=0.9), the density of X-ray luminous ($L_X > 10^{42}$ erg s$^{-1}$) AGN is found to increase (e.g., Ananna et al. 2019). How does the X-ray luminosity of the $\sim10^7$ K gas vary as a function of z? Here we attempt to address this question observationally for the first time, by exploring the redshift evolution of $L_{X,GAS}$. Because the direct relation between $L_{X,GAS}$ and z is biased by the limiting sensitivity of the observations (i.e., Malmquist Bias), we use X-ray luminosity functions (XLFs) in this study.

This work supersedes previous studies of galaxy XLF (e.g., Kim et al. 2006, Ptak et al. 2007, and Tzanavaris & Georgantopoulos 2008) in two ways. First, our sample of ~600 galaxies extracted from the Chandra Galaxy Catalog (Kim et al., 2023a) is much larger than those used in previous work, and thus it gives a far better determination of the XLF of the total emission of normal galaxies. Second, at difference from previous galaxy XLF works, we also estimate $L_{X,GAS}$ for each galaxy, instead of using the total X-ray luminosity, which also includes the contribution from X-ray binaries and possibly hidden nuclear AGNs. Because the hot gas emission peaks at lower energies than that of AGNs and X-ray binaries (e.g., Kim et al. 1992), we primarily use the 'soft' S energy band (0.5-1.2 keV) from the Chandra Source Catalog (CSC2.0, Evans et al. 2010). The X-ray emission of a hot gas with kT ~ 1 keV peaks in the S-band (0.5 – 1.2 keV) at z = 0. Because the peak is redshifted out at higher z, we can explore the hot gas out to z ~ 0.6.

This paper is organized as follows. In Section 2, we describe our sample selection. In Section 3, we present the sample properties and discuss the K-corrections for optical and X-ray data. In Section 4, we describe the determination of the hot gas X-ray luminosity. In Section 5, we present both a new XLF of galaxies (inclusive of all the emission components) and the hot gas XLF. We discuss the dependence of the XLF with z, and other implications in Section 6. Our results and conclusions are summarized in Section 7.

Throughout the paper, we adopt the following cosmological parameters from Bennett et al. (2014): Ho = 69.6 km $s^{-1}$ / Mpc, $\Omega_M$ = 0.286, and $\Omega_\Lambda$= 0.714.

## 2. THE SAMPLE

Our sample is extracted from the Chandra Galaxy Catalog (CGC; Kim et al. 2023a), which was derived by cross-matching the X-ray sources from the Chandra Source Catalog version 2[1] (CSC2.0) with the optical surveys SDSS DR16[2], PanSTARRS DR2[3], and Legacy DR8[4]. The CGC consists of 8547 galaxies of which 24% are spectroscopically identified in SDSS as galaxies. Additionally, Kim et al. (2023a) identified galaxies with: (1) the joint selection of $L_X$ and the X-ray to optical flux ratio $F_{XO}$ [ $L_X/4.64 \times 10^{42} < (F_{XO} / 0.1)^{-1/3}$ ], where the X-ray flux/luminosity is in the 0.5-7 keV band, the optical mag in the r-band and log ($F_{XO}$) = log ($F_X$) + 5.31 + r /2.5; and (2) the joint selection of $L_X$ and the WISE W1-W2 color (Wright et al. 2010) [ log ($L_X$) < -2.2 (W1-W2) + 43.78 ]. These selection criteria exclude X-ray sources with QSO counterparts since these have higher $L_X$, $F_{XO}$, and W1-W2 (e.g., Maccacaro et al. 1988; Stern et al. 2012). Kim et al. (2023a) estimates that the QSO contamination of the CGC is ~ 5%, comparable to the false match probability in cross matching the CSC2.0 and optical catalogs. Counting at the 5th and 95th

[1] https://cxc.harvard.edu/csc/
[2] https://www.sdss.org/dr16
[3] https//panstarrs.stsci.edu
[4] https://www.legacysurvey.org/

percentiles, the redshift range of the CGC sample is z = 0.04 – 0.7; the (0.5 – 7 keV) X-ray luminosity range is $L_X = 2 \times 10^{40} – 2 \times 10^{43}$ erg $s^{-1}$. The CGC includes obscured or diluted AGNs at the high $L_X$ end ($>10^{42}$ erg $s^{-1}$), even though they are spectroscopically classified as galaxies (see Kim et al. 2023b for those X-ray bright optically normal galaxies, XBONG). We will discuss the possible AGN contamination in Section 6.1

The sample for our XLF study was extracted from five Chandra large-area surveys inside the SDSS footprint. They are outside the Galactic plane and away from any known complex objects, hence optimal for the XLF study of normal galaxies. In Table 1, we list the approximate central position, the number of CGC galaxies, the sky coverage area, the range of exposures, and the main references of each survey. Note that the Lockman Hole field consists of three non-overlapping sub-fields, including the SWIRE field (see the references in Table 1). The total number of galaxies from these five surveys is 653. Our sample is considerably larger than any samples previously used for XLF studies of normal galaxies. We do not attempt to separate early and late-type galaxies in this study because of the lack of an accurate classification available. A simple discrimination based on color is discussed in Appendix A, where we find that the XLFs of red and blue galaxies are consistent with the joint XLF discussed in this paper.

Of 653 galaxies, 265 (41%) have SDSS spectral data. The other galaxies have photo-z from SDSS (50%), PanSTARRS (7%), and Legacy (2%). As shown in Kim et al. (2023a), the photo-z and spec-z in CGC are consistent with an outlier fraction of ~5% when determined within 0.2 dex by comparing with spec-z. The outliers are mostly found at the lowest and highest redshifts, which are beyond our interest range.  In the mid-range of z = 0.03-0.6, as shown in Figure C1 (Kim et al. 2023a), the outlier fraction is 4%. See Appendix B for a discussion of the redshift uncertainty.

Table 1. Selected Chandra deep survey fields

| field name | (ra, dec) deg | # of CGC galaxies | sky coverage $deg^2$ | exposure ksec | ref |
|---|---|---|---|---|---|
| Bootes | 218, 34 | 214 | 8.9 | 5-100 | 1 |
| COSMOS | 150, 2 | 210 | 2.1 | 5-50 | 2 |
| Lockman Hole (SWIRE) | 160, 58 | 104 | 2.7 | 5-80 | 3 |
| DEEP2 Field 3 (Strip 82) | 352, 0 | 22 | 1.2 | 10 | 4 |
| DEEP2 Field 1 (AEGIS) | 215, 53 | 103 | 1.1 | 200-800 | 5 |
| Total | | 653 | 16 | | |

References to the Chandra observations
1. Masini et al. 2020; 2. Civano et al. 2016, Marchesi et al. 2016; 3. Kenter et al. 2003; Yang et al. 2004, Polletta et al. 2006 (SWIRE); 4. Goulding et al. 2012; 5. Laird et al. 2009, Nandra et al. 2015

In Figure 1, we show the r magnitude distribution of our sample. Each sub-sample is marked with a different color, as indicated in the figure. The SDSS spec-z sample (cyan) consists of two sub-groups, one at r < 18 mag and another at r = 18-21 mag. r ~18 mag was the magnitude limit originally set for the targets in the early SDSS spectroscopic survey[5]. For r < 18 mag, 83% of galaxies have spec-z, but for r > 18 mag only 30% (152 / 516) of the galaxies have spec-z. The SDSS photometric sample (green) with a survey limit at r ~ 22 mag[6] includes most galaxies with r = 18 – 22 mag. The PanSTARRS (grey) and Legacy (magenta) photometric samples go deeper to r ~ 23 mag. Although spec-z is more reliable than photo-z, we need to use both to build the luminosity function and avoid incompleteness.

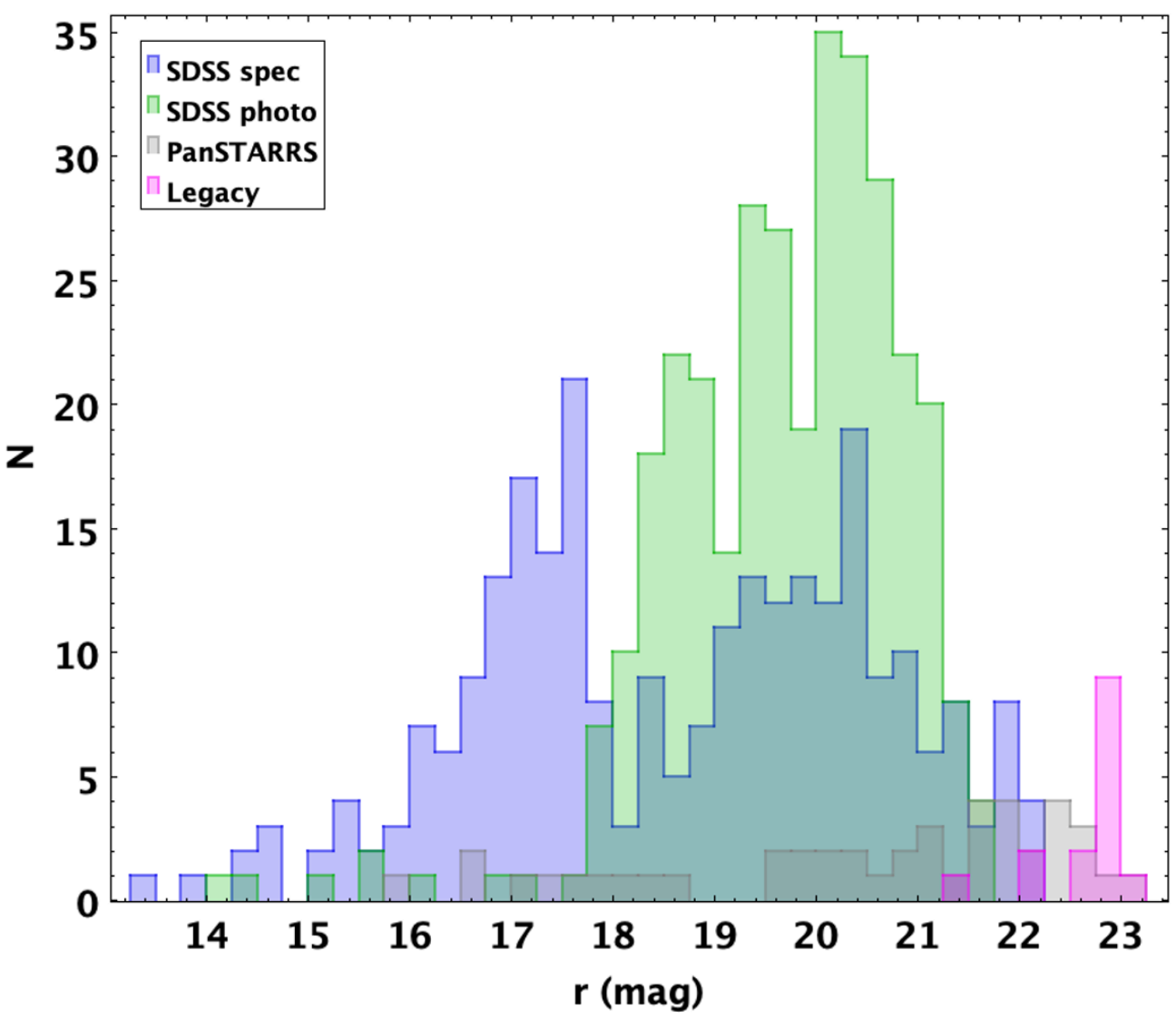

Figure 1. The histogram of r magnitude for the SDSS spec-z (cyan), SDSS photo-z (green), PanSTARRS (grey), and Legacy (magenta) subsamples.

## 3. SAMPLE PROPERTIES

### 3.1 K-corrected Optical Properties

We use the Python package, *kcorrect*[7] (Blanton & Roweis 2007) for evaluating the optical K-correction. This tool utilizes multi-band optical magnitudes instead of relying on a single color. We take ugriz for SDSS, griz for PS, and grz for Legacy. We convert all the magnitudes to the AB mag system. We adopt the recipes in Blanton & Roweis (2007) to convert the SDSS Asinh mag[8] to the AB system, Tonry et al. (2012) to convert between the PanSTARRS and SDSS magnitude systems, and the Legacy website[9] to convert between the Legacy and PanSTARRS magnitude systems. We compared the K-correction with those evaluated with the tool *kcor* (Chilingarian et al. 2010) and confirmed that they are consistent at z < 0.6 within a few tens of magnitude.

[5] https://classic.sdss.org/dr7/algorithms/target.php
[6] https://www.sdss4.org/dr17/imaging/other_info/#DepthsoftheSDSSphotometricsurvey
[7] https://kcorrect.readthedocs.io/en/5.1.1/index.html
[8] https://www.sdss4.org/dr17/algorithms/magnitudes/#SDSSAsinhMagnitudes
[9] https://www.legacysurvey.org/dr10/description/#photometry

We also use *kcorrect* to determine the stellar mass. With multiband photometry, this tool provides more reliable measurements than the method using a single color. Comparing with the stellar masses estimated with single color recipe in Bell et al. (2003), *kcorrect* provides a systematically lower stellar masses, the mean ratio being 0.6. A similar discrepancy was reported by Lehmer et al. (2016). We use the stellar mass to calculate the expected X-ray binary contribution with the scaling relations in Lehmer et al. (Section 4). The errors in our results due to the uncertainties in stellar mass are insignificant because we are applying the scaling relations, which were already calibrated with a similar stellar mass.

In Figure 2, we plot the k-corrected optical luminosity $L_r$ and stellar mass $M_{STAR}$ as functions of z, for the spec-z (red dots) and photo-z (blue dots) samples. $L_r$ does not show a strong redshift dependency in the parameter space outlined by $L_r = 1 \times 10^{10} - 2 \times 10^{11}$ $Lr_{\odot}$ (or $M_{STAR} = 2 \times 10^{10} - 3 \times 10^{11}$ $M_{\odot}$) and z < 0.6, which is uniformly filled with galaxies. Exceptions are at the lowest and highest z. Low luminosity/mass galaxies ($L_r = 10^9$ - $10^{10}$ $Lr_{\odot}$) are only visible at z < 0.2. At z > 0.6, luminous (massive) galaxies are preferentially included. Given the optical selection effect (and the X-ray limitation described in section 3.2), we explore the XLF only at z < ~0.6.

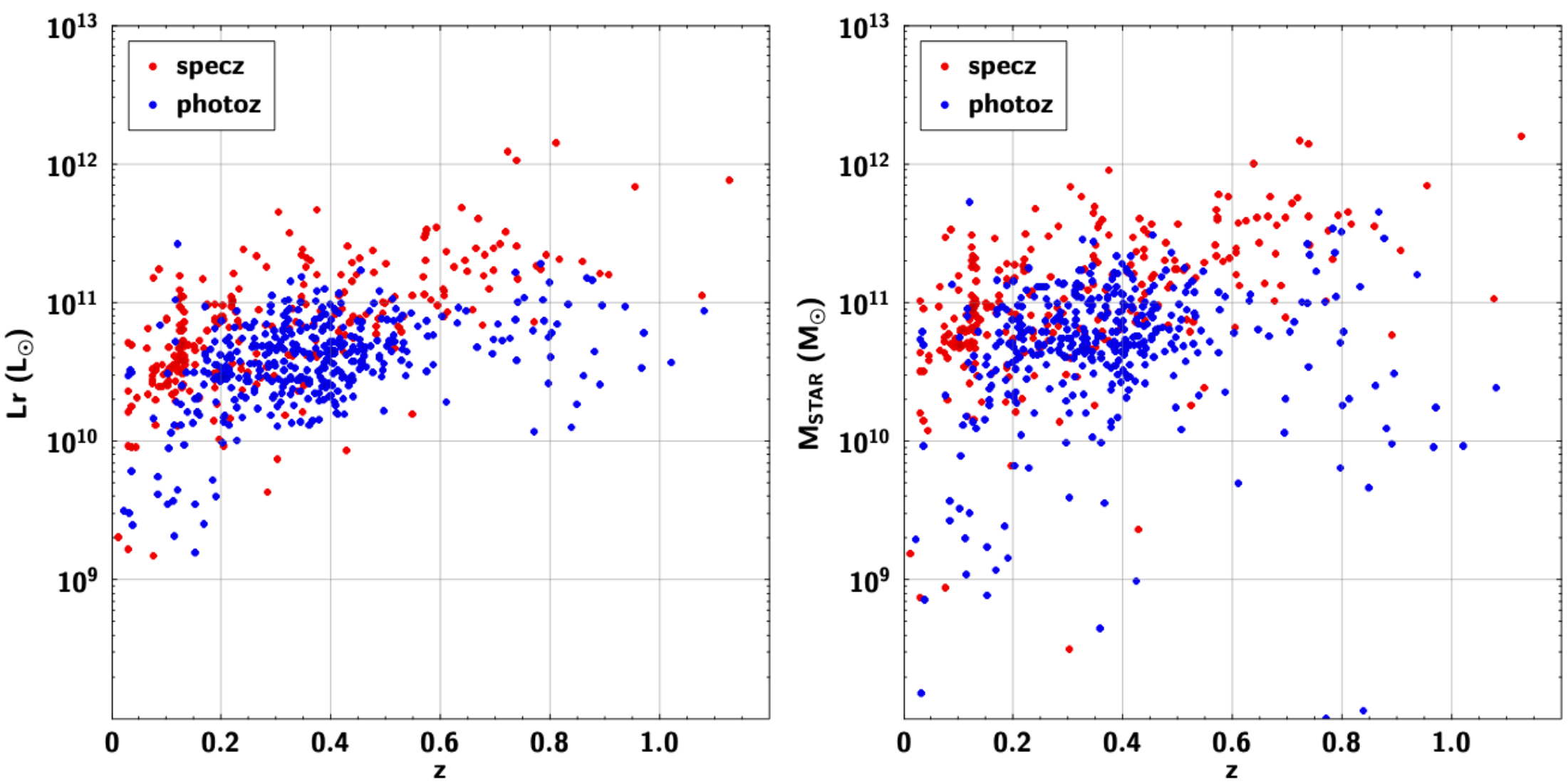

Figure 2. (left) The optical luminosity (in solar units, after K-correction) in the r-band and (right) the stellar mass is plotted against redshift. The galaxies with spec-z and photo-z are red and blue points, respectively.

## 3.2 K-corrected X-ray Properties

The X-ray emission of normal, non-AGN, galaxies is dominated by the emission of hot gas and populations of X-ray binaries, including low-mass (LMXBs) and high-mass X-ray binaries (HMXBs) (Fabbiano 1989). Unlike nearby galaxies, where we can detect X-ray point sources and separate them from the diffuse emission (e.g., Kim et al. 2019), for our sample galaxies, we can only separate the two emission components in the X-ray spectra or using the scaling laws derived from Chandra observations of X-ray binary populations.

The thermal X-ray emission from the hot gas is soft and peaks at ~0.8 keV, while the power-law component from the X-ray binaries is harder. This is illustrated in the left panel of Figure 3, for a typical early-type (T = -5) galaxy NGC 720, at a distance D = 28 Mpc, with $L_X$ (0.5-1.2 keV) ~ 5 x $10^{40}$ erg $s^{-1}$ and $M_{STAR} \sim 10^{11}\ M_\odot$. In the soft band of the CSC2.0, S-band=0.5 - 1.2 keV, the thermal component from the hot gas dominates, while in the hard band, H-band=2.0 – 7.0 keV, the power-law component from the X-ray binaries dominates. The two components are mixed in the medium energy M-band=1.2-2.0 keV. The AGN component, if non-negligible, is also hard and can be included in the power-law model. In the S-band, the expected X-ray contribution from the X-ray binaries is relatively small and the X-ray luminosity from the hot gas can be reliably measured with a small correction (see Section 4). The power-law component can be measured in the H-band with little contamination from the hot gas. In the case of NGC 720, LMXBs dominate the hard band because of the old stellar population of this galaxy and its lack of strong AGN emission.

At higher z, the effect of red-shifting it to make the gas component less prominent in the S-band and the power-law contribution larger, therefore a correction for the power-law component is important at higher z (Figure 3, right panel; see section 4). This practically sets the z limit for exploring the hot gas XLF of galaxies at z ~ 0.6.

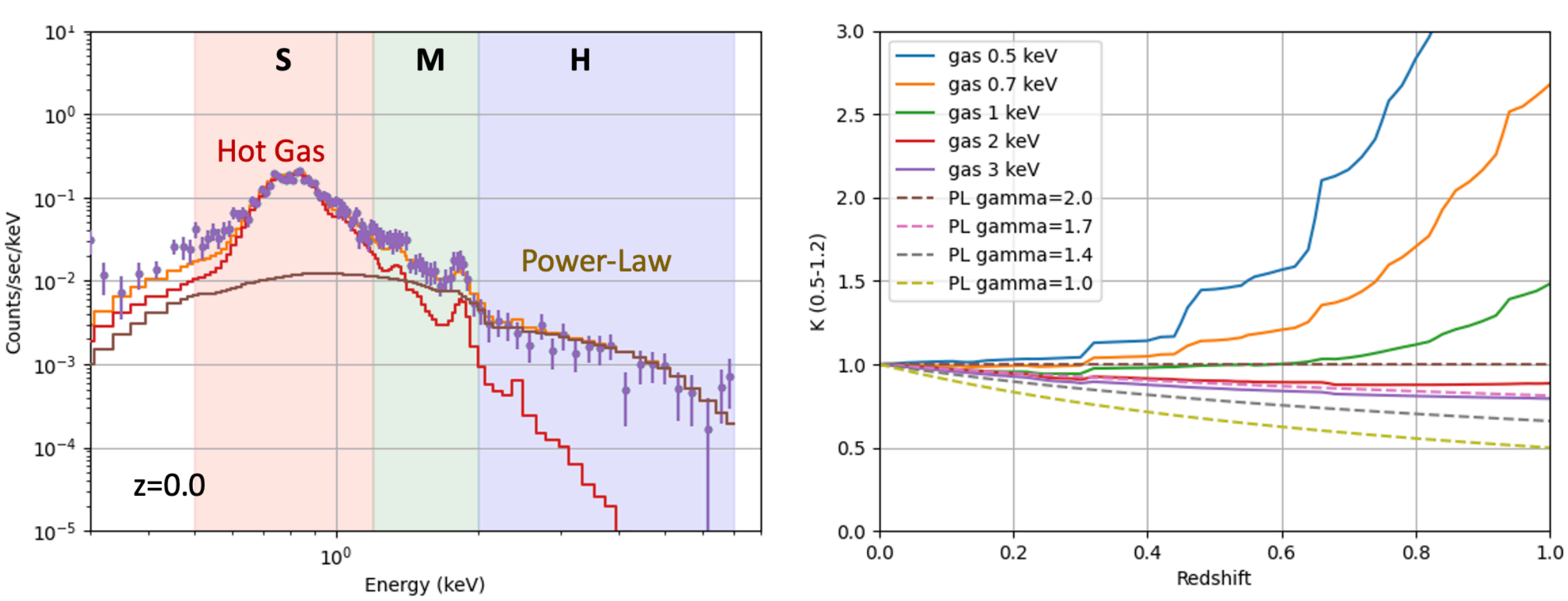

Figure 3. (left) The Chandra X-ray spectrum of a typical local early-type galaxy, NGC 720 with $L_X$ (0.5-1.2 keV) ~ 5 x $10^{40}$ erg $s^{-1}$. In the S-band (0.5 - 1.2 keV), the thermal APEC[10] component from hot gas (red color; kT ~ 0.6 keV) dominates, while in the H-band (2 - 7 keV), the power-law (PL) component (brown color; a photon index of 1.4) dominates. (right) X-ray K-corrections in the S-band (0.5-1.2 keV) for the APEC (gas) and PL components with multiple gas temperatures and PL indices.

[10] https://cxc.cfa.harvard.edu/sherpa/ahelp/xsapec.html

Since the two X-ray emission components, hot gas and power-law, have distinct spectra, we must apply the K-correction to each component separately. We use *calc_kcorr*[11] in the *Sherpa*[12] package to estimate the X-ray K-correction. We use the APEC model for the hot gas emission. We set the absorption column density, $N_H$=3 x $10^{20}$ $cm^{-2}$, which is appropriate for a high galactic latitude, as in our sample. We set the abundance to be solar, but the results do not change with a lower abundance, e.g., ~0.3 solar, which may be appropriate for high-z galaxies. In the right panel of Figure 3, we show the X-ray K-corrections in the S-band, defined as the ratio of the flux in the rest-frame energy band to that in the observed band. The K-correction for the APEC component is > ~1 (for kT < 3 keV) while the K-correction for the power-law component (dashed lines) is < 1 (for photon index < 2). In both cases, the correction is smaller than the typical optical K-correction.

In Figure 4 (left panel), we show the corrected X-ray luminosity in the (0.5-1.2 keV) S-band as a function of z. The K-correction of the APEC model was applied to $L_X$(S), assuming the hot gas dominates in the S-band. To estimate the gas temperature, we used the relation between $L_{X,GAS}$ and $T_{GAS}$ from Kim & Fabbiano (2015) after converting $L_{X,GAS}$ from the energy band 0.5-1.2 keV to the 0.3-8.0 keV band used in that paper.

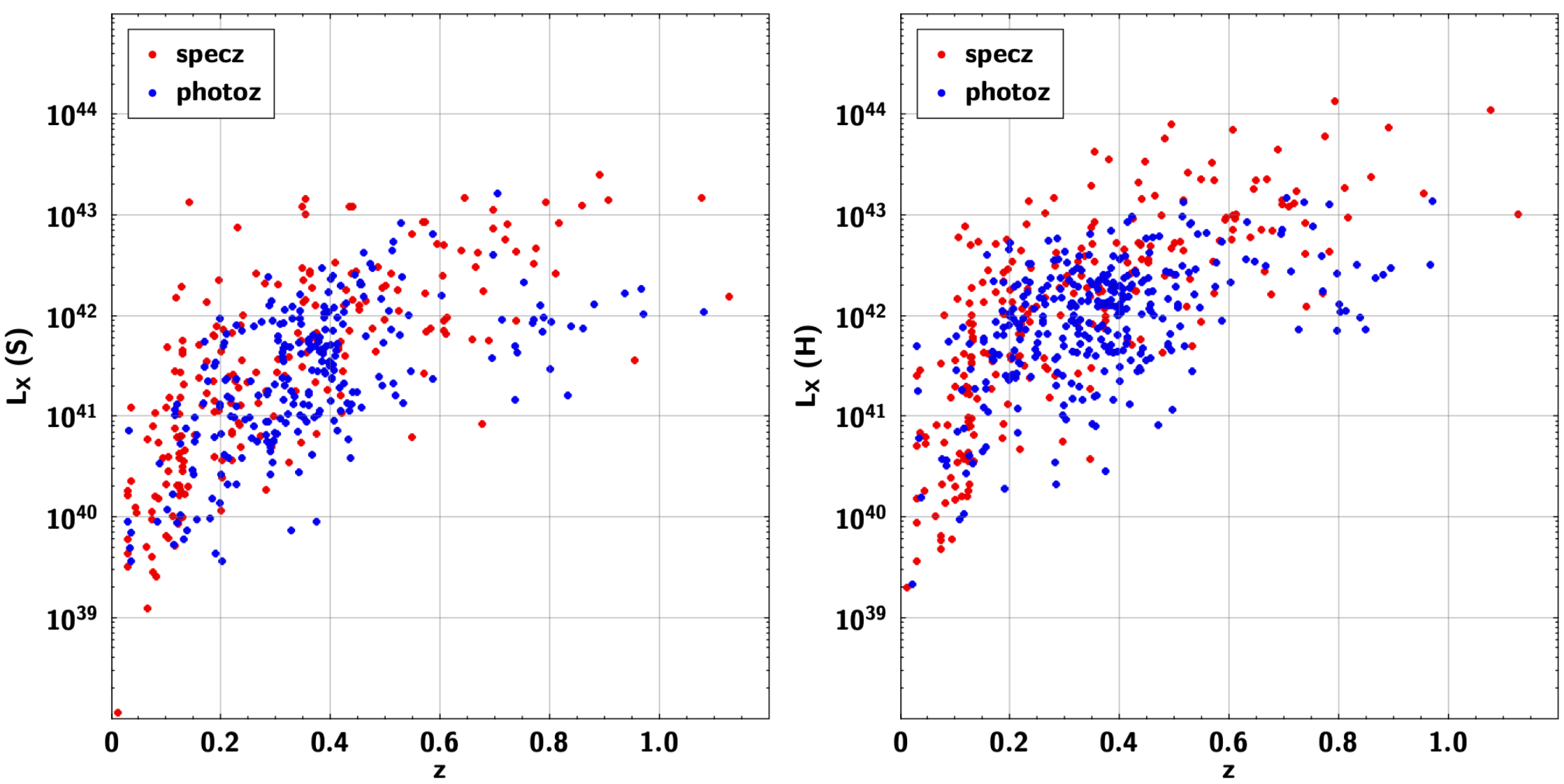

Figure 4. The X-ray luminosity (left) in the S-band (0.5-1.2 keV) and (right) H-band (2-7 keV) plotted against redshift. The K-correction for the APEC (power-law) component is applied for the S (H) band. The galaxies with spec-z and photo-z are red and blue points, respectively.

In contrast to the optical luminosity – z plot (Figure 2), a redshift trend is visible in the entire z range in the $L_X$(S) – z plot, i.e., the X-ray detection limit dictates the minimum $L_X$ as a function of z in the sample. We will consider this incompleteness when building the XLF (Section 5). The galaxies with spec-z and photo-z are marked separately in Figure 4. They are well mixed at z <

[11] https://cxc.cfa.harvard.edu/sherpa/threads/calc_kcorr/
[12] https://cxc.cfa.harvard.edu/sherpa4.16/

0.6. At higher z, most galaxies with photo-z lie at $L_X < 3 \times 10^{42}$ erg s$^{-1}$. They are from the PanSTARRS and Legacy surveys, which go deeper than the SDSS (See Figure 1), indicating that the optical survey may be incomplete at higher z.

Similarly, the X-ray luminosity in the H-band (2.0-7.0 keV) in the right panel of Figure 4 is corrected by the K-correction of the power-law model, assuming the hot gas component is negligible in the H-band. We take the power-law photon index of 1.4 (see Section 4 for this choice).

## 4. THE X-RAY LUMINOSITY OF THE HOT GAS

The hot gas of normal galaxies is the primary X-ray source in the S-band, as seen in Figure 3. In first approximation, we may equate $L_{X,GAS}$ to the S-band luminosity: $L_{X,GAS}$ (S) = $L_X$ (S). However, $L_X$(S) may still include sizeable X-ray emission from X-ray binaries (LMXBs and HMXBs) in the gas-poor ($L_X < 10^{40}$ erg s$^{-1}$) galaxies. Obscured/diluted AGNs may contribute in luminous or high-z galaxies (Kim et al. 2013b). To accurately measure the X-ray luminosity of the hot gas, $L_{X,GAS}$, we apply further corrections to exclude these other X-ray emission components.

### 4.1 Subtraction of the X-ray Binary Contribution using scaling laws

It is well-known that the X-ray luminosity of the population of X-ray binaries is closely related to the stellar property of galaxies, i.e., stellar mass and SFR: the $L_X$ from the population of LMXBs ($L_{X,LMXB}$) in a given galaxy is linearly proportional to $L_K$ and $M_{STAR}$ (e.g., Boroson et al. 2011); instead, the $L_X$ from populations of HMXBs ($L_{X,HMXB}$) is closely related to the SFR (Mineo et al. 2012). Both $L_{X,LMXB}$ and $L_{X,HMXB}$ increase with increasing redshift because the SFR peaks at z = 2-3 (Fragos et al. 2013; Lehmer et al. 2016).

We estimate the expected $L_{X,LMXB}$ by applying the scaling relations as a function of $M_{STAR}$ and redshift (Lehmer et al. 2016). To convert the luminosities in different energy bands from 2.0-10.0 keV to 0.5-1.2 keV, we take a power-law model with a photon index of 1.4 and the line-of-sight Galactic hydrogen column density, $N_H = 3 \times 10^{20}$ cm$^{-2}$. The expected $L_{X,LMXB}$ in the S-band is compared with the observed $L_X$(S) in the left panel of Figure 5. The diagonal line, indicating $L_X$(S) = $L_{X,LMXB}$(S), lies at the low end of the observed $L_X$(S). The LMXB contribution is significant only for low luminosity ($L_X$ < a few x $10^{40}$ erg s$^{-1}$), local (z < 0.2) galaxies.

Similarly, we estimate the expected $L_{X,HMXB}$ as a function of star formation rate (SFR) and redshift, with the scaling relations given by Lehmer et al. (2016). To convert the luminosities from 2.0-10.0 keV to 0.5-1.2 keV, we take a power-law model with a photon index of 1.4 and $N_H = 3 \times 10^{20}$ cm$^{-2}$. To determine SFR, we use the relation between SFR and $M_{STAR}$ for the star-forming spiral galaxies, the so-called galaxy main sequence (Speagle et al. 2014). The expected $L_{X,HMXB}$ in the S-band is compared with the observed $L_X$(S) in the right panel of Figure 5. Similar to the LMXBs, $L_{X,HMXB}$ in the S-band is lower than $L_X$(S). We note that $L_{X,HMXB}$ is overestimated in early-type galaxies because the relation is only applicable to star forming late-type galaxies. Given that the HMXB contribution is not significant in most galaxies, we apply the same $L_{X,HMXB}$ regardless of the galaxy type. If there is no significant AGN contribution, the hot gas luminosity in the S-band is

Method A: $L_{X,GAS}$ (S) = $L_X$ (S) - $L_{X,LMXB}$ ($M_{STAR}$, z) - $L_{X,HMXB}$ (SFR, z)

where $L_X$ (S) is the X-ray luminosity in the S-band K-corrected for the APEC model.

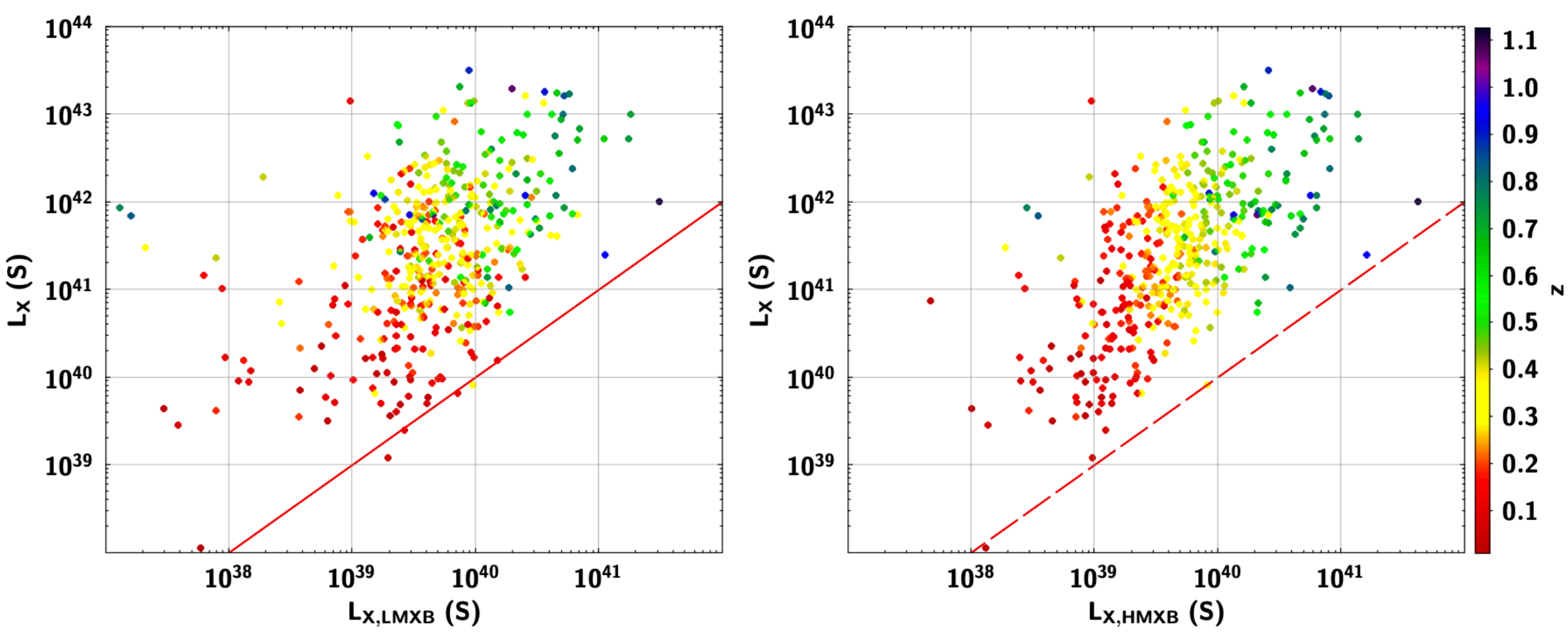

Figure 5. The observed $L_X$ in the S-band is compared with (left, red line) the expected $L_{X,LMXB}$, and (right, dashed red line) the expected $L_{X,HMXB}$. The data points are color-coded by z.

4.2 Subtraction of the hard spectral component, which should account for both XRBs and AGNs

As described in Section 2, our galaxy sample does not include AGNs with known optical emission lines. Moreover, AGNs are also excluded by the $L_X$ – $F_{XO}$ or $L_X$- $W_{12}$ selection (Kim et al. 2023a). However, some heavily obscured AGNs or AGNs diluted by the stellar light of the host galaxy may be included (Kim et al. 2023b). We assume that a hard power-law model represents both the AGN component and the LMXB and HMXB emission, which have similar hard X-ray spectra, and that the X-ray emission in the H-band is entirely from this power-law component with a negligible hot gas contribution. We estimate the power-law component in the S-band from $L_X$(H), assuming a power-law model with a photon index of 1.4 (see below in Section 4.3 for this choice of power-law index) and line-of-sight Galactic absorption ($N_H$ = 3 x $10^{20}$ $cm^{-2}$) with no intrinsic absorption. This gives a second estimate of the hot gas luminosity:

Method B: $L_{X,GAS}$ (S) = $L_X$ (S) – $L_X$ (H) * C

where $L_X$ (S) and Lx (H) are the X-ray luminosities in the S-band and H-band K-corrected for the APEC and power-law model, respectively. C is the conversion factor of the power-law component from the H-band to the S-band.

4.3 Comparison of the two estimates of $L_{X,GAS}$

Considering the 453 galaxies detected in the S-band, Method A gives positive (> 0) values of $L_{X,GAS}$ for 446 (98%) galaxies, and Method B for 292 (65%) galaxies. See more discussions on the excluded galaxies with negative $L_{X,GAS}$ in Section 6.1. Since Method A only subtracts the contribution of the X-ray binary population, it will over-estimate $L_{X,GAS}$, if a low-luminosity AGN is present. Method B in principle should give a more rigorous estimate but there are some issues. If the AGN is intrinsically absorbed, with $N_H > 10^{22}$ cm$^{-2}$, Method B will over-correct the power-law component in the S-band. Also, the hot gas emission may not be negligible in the H-band, particularly in the hot gas-rich galaxies at low z, and this will increase the estimated power-law component in the S-band. In both cases, Method B will tend to under-estimated $L_{X,GAS}$. We conclude that the two methods set lower and upper limits of $L_{X,GAS}$. In Figure 6, we compare $L_{X,GAS}$ measured with the two methods. The two estimates are well correlated and differ at most by a factor of 3 at $L_{X,GAS} > 10^{40}$ erg s$^{-1}$.

Throughout this work, we have set the power-law photon index to 1.4, which is slightly lower than the canonical value of 1.7 for unobscured AGNs and X-ray binaries. This is to mitigate the potential over-correction of non-gas components in Method B. Note that the index 1.4 is still too large for obscured AGNs because a power-law index 1.7 with $N_H = 10^{22}$ cm$^{-2}$ corresponds to a power-law index < 1 with normal absorption ($N_H = 3 \times 10^{20}$ cm$^{-2}$). Also, the photon indices 1.4 and 1.7 do not make any significant differences in the K-correction (Figure 3) and in the estimated $L_{X,GAS}$ using Method A. (Section 4.1).

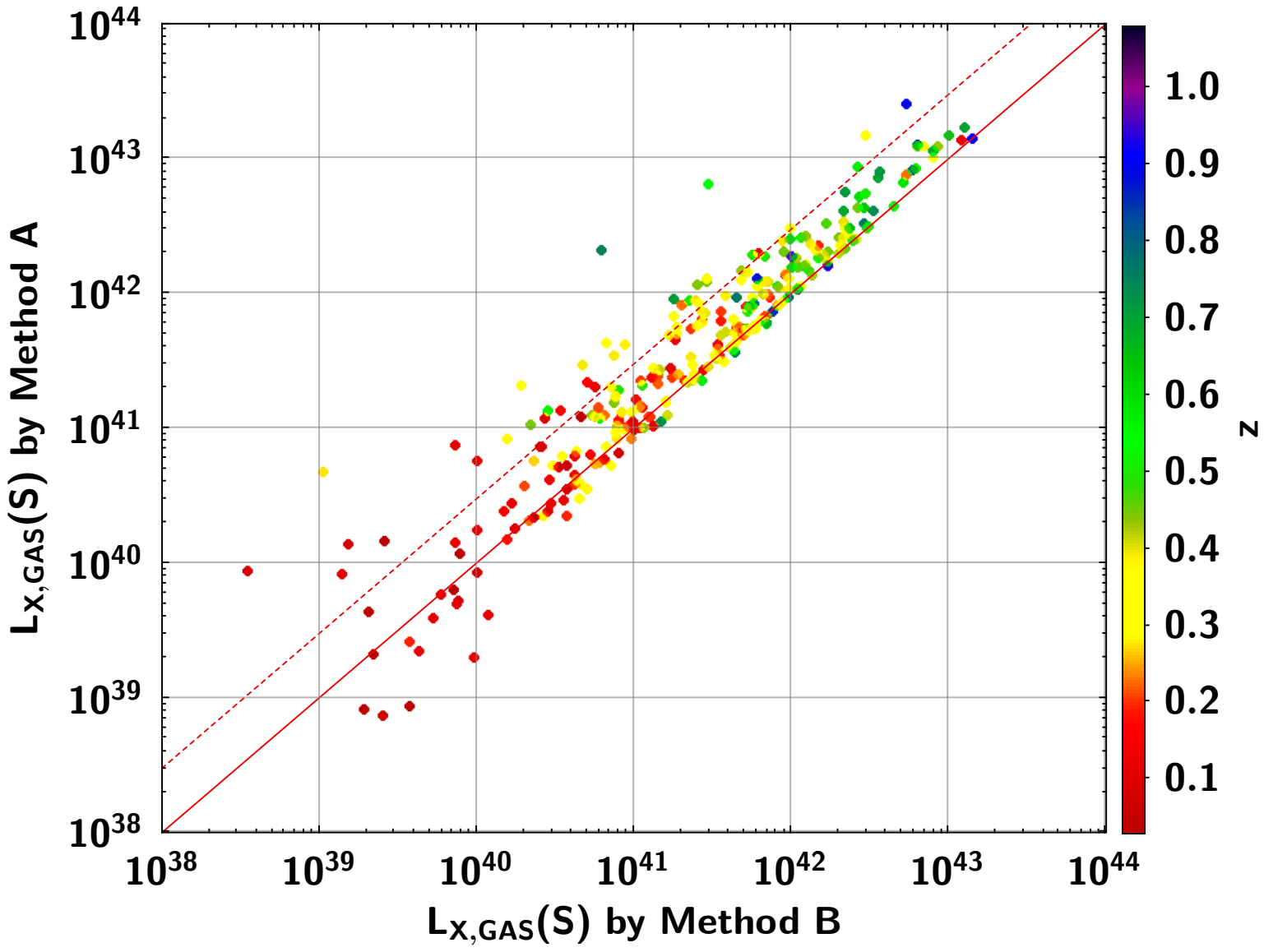

Figure 6. Comparison of $L_{X,GAS}$ estimated by two methods. The solid and dotted lines indicate y=x and y=3x, respectively.

## 5. THE HOT GAS X-RAY LUMINOSITY FUNCTION

Using the $1/V_a$ method (Schmidt 1968), the source density in a given luminosity bin is given by

$$\Phi(L) = \sum_{i=1}^{N} \frac{1}{V_{a,i}}$$

where $N$ is the number of galaxies in a given luminosity bin and $V_{a,i}$ is the accessible volume for galaxy $i$. Following Hogg (1999), the co-moving volume can be written as

$$V_{a,i} = \frac{c\,\Omega_i(fx)}{H_o} \int_{Z\min}^{Z\max} \frac{(1+z)^2 D_{A,i}^2}{(\Omega_M (1+z)^3 + \Omega_A)^{1/2}} \quad dz$$

where $\Omega_i(f_X)$ is the sky coverage for the X-ray flux ($f_X$) of galaxy i, $D_{A,i}$ is the angular diameter distance at redshift z, and $z_{min}$ and $z_{max}$ are the minimum and maximum redshifts possible for a source to stay in the luminosity bin.

Because the X-ray detection is flux-limited, the sky coverage area varies as a function of $f_X$ in each energy band. We take the sensitivity map[13] per energy band, which is available in the CSC2.0 database, and apply the science thread recipe[14] to calculate the sky coverage as a function of $f_X$. The sky coverage for our sample is given in Figure 7. In the S-band, the total area for flux > $10^{-14}$ erg $s^{-1}$ $cm^{-2}$ is ~16 $deg^2$, and the limiting flux is a few x $10^{-16}$ erg $s^{-1}$ $cm^{-2}$.

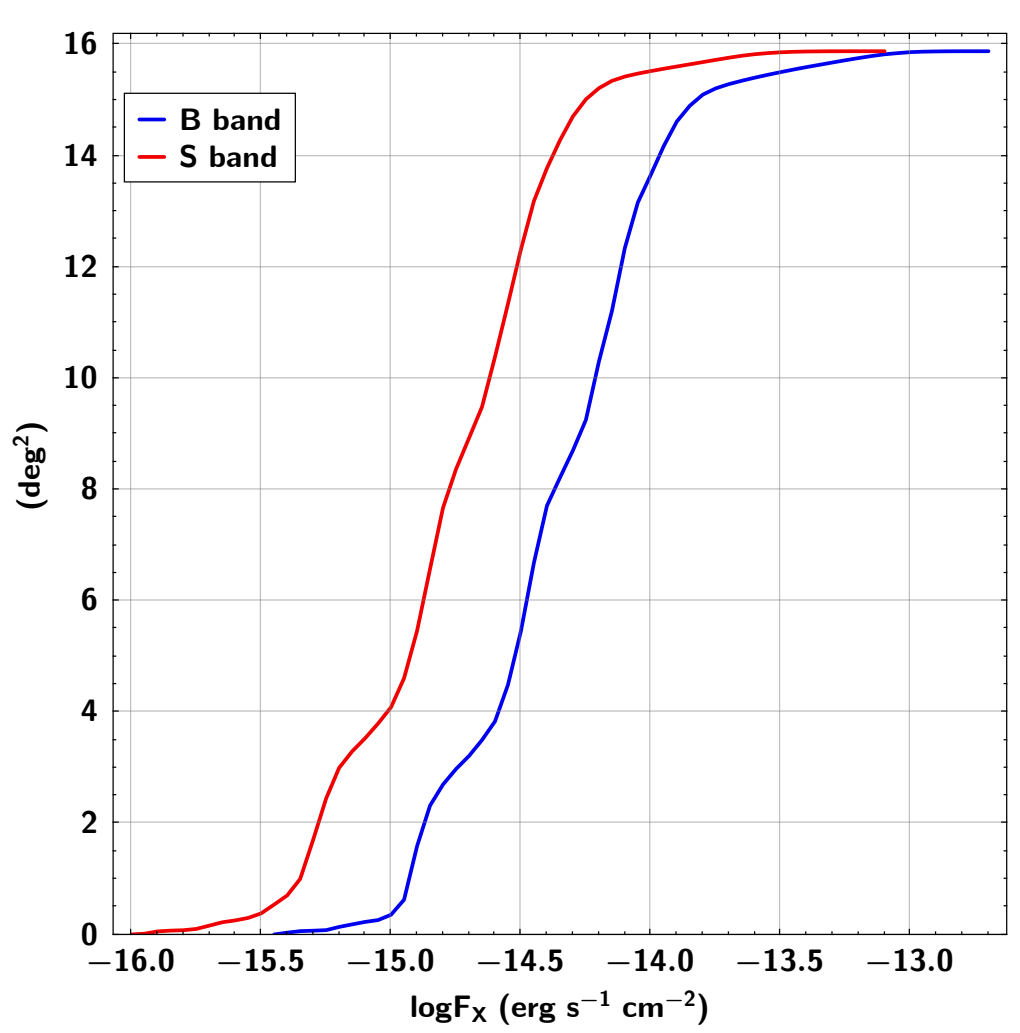

Figure 7. The sky coverage area is plotted as a function of the X-ray flux for the S (0.5-1.2 keV) and B (0.5 – 7 keV) energy bands.

First, we build the Chandra broad B-band (0.5-7.0 keV) XLF and compare it with those in the literature in the same band. Out of 653 galaxies detected in the B-band, we select the 575 galaxies within z < 0.6, which as we discussed earlier is the practical limit for reliably building the XLF. The B-band luminosity includes all the emission sources from the entire galaxy - hot gas, LMXBs,

[13] https://cxc.harvard.edu/csc/data_products/stack/sens3.html
[14] https://cxc.cfa.harvard.edu/csc/threads/cumulative_sky_coverage/notebook.html

HMBXs, and possibly an unidentified AGN. In Figure 8, the XLF from this work (red points) is compared with the previously published results with the CDF galaxies (cyan; Norman et al. 2004), the NHS galaxies (green; Georgantopoulos et al. 2005), and the ChaMP galaxies (magenta; Kim et al. 2006). The XLFs are consistent with each other within the error in the range $L_X(B) = 10^{39}$ - $10^{42}$ erg s$^{-1}$. Galaxies with $L_X > 10^{42}$ erg s$^{-1}$ are XBONGs (Kim et al. 2023b; see Section 6.1). The statistical error of our XLF is the smallest, given that we use the largest number of galaxies. For visibility and comparison, we plot a line with a slope of -1 in all XLF plots.

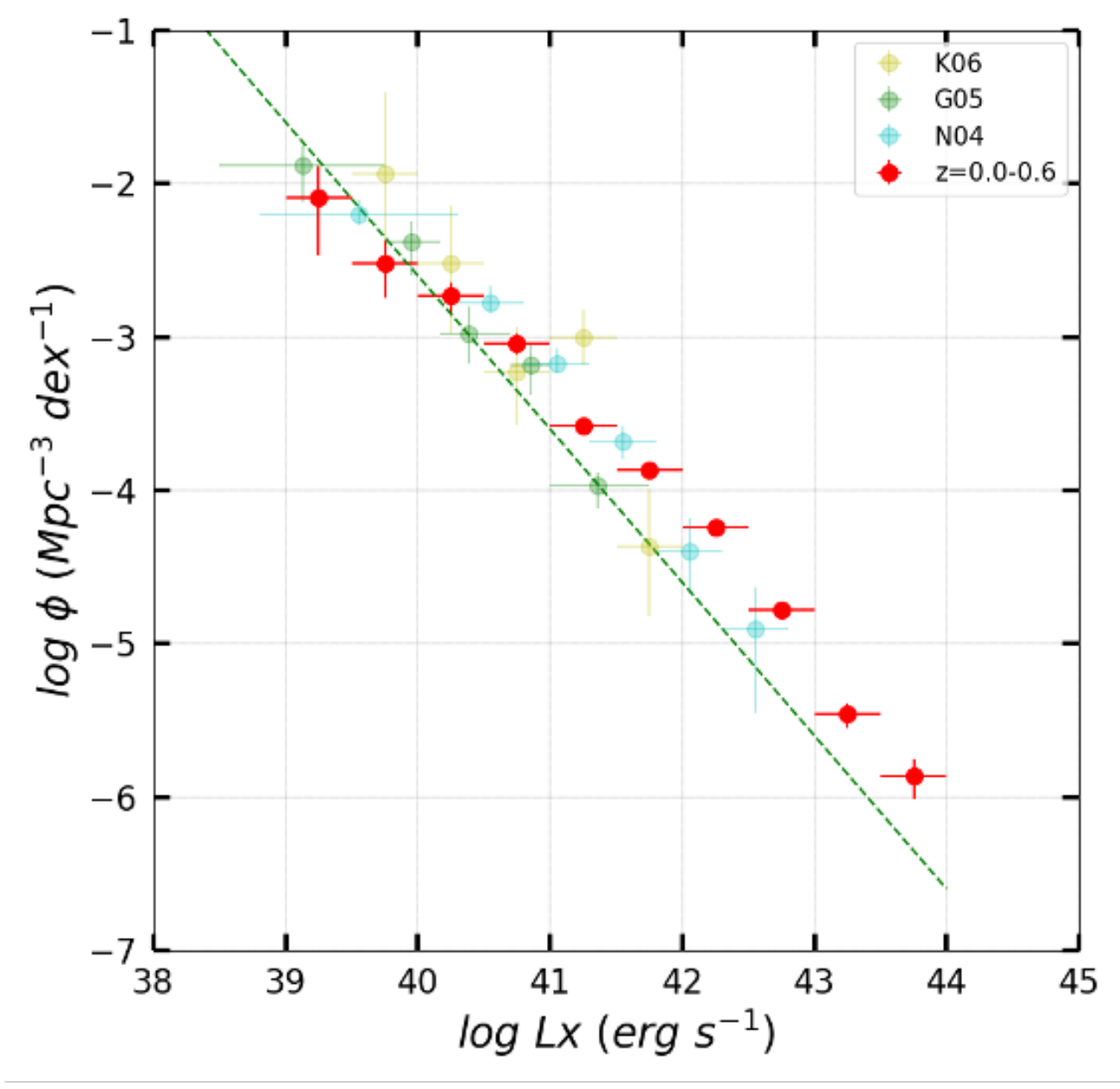

Figure 8. The B-band (0.5-7.0 keV) XLF of galaxies. Also plotted are previously measured XLFs by Norman et al. (2004), Georgantopoulos et al. (2005), and Kim et al. (2006). The dashed line with a slope of -1 ($\log \phi = -1 * \log L_X + 37.4$) is for visibility. The same line is drawn in all XLF figures below. Note that the X-ray luminosity includes all X-ray emission components (AGN, X-ray binaries, and hot gas).

Second, we build the soft-band XLF with $L_X(S)$ (the left panel of Figure 9). Out of 453 galaxies detected in the S-band, 398 galaxies have $z < 0.6$. We use the S-band sky area function in Figure 7. The S-band XLF is shifted, when compared with the B-band XLF, to the left [because $L_X(S) < L_X(B)$ ] and to the lower numbers (because some galaxies are not detected in the S-band). The diagonal line with a slope of -1 matches the data points well in the luminosity range $L_X(S) = 10^{40}$ - $10^{42}$ erg s$^{-1}$, where our XLF is most reliable. At the low luminosity end, $L_X < 10^{40}$ erg s$^{-1}$, the uncertainty is large, and our sample is possibly incomplete in the optical band for which we apply no correction. At the high luminosity end, $L_X > 10^{42}$ erg s$^{-1}$, some sources may host AGNs (see Section 6).

We split the XLF into three redshift bins in the right panel of Figure 9: z = 0.1 (0.0-0.2), z = 0.3 (0.2-0.4), and z = 0.5 (0.4-0.6). The number of galaxies (per unit co-moving volume) increases by a factor of 2 - 6 from z = 0.1 to z = 0.3 at $L_X(S) = 10^{40}$ - $10^{41.5}$ erg s$^{-1}$. Similarly, the number of galaxies increases by a factor of 3 - 5 from z = 0.3 to z = 0.5 at $L_X(S) = 10^{40.5}$ - $10^{41.5}$ erg s$^{-1}$.

Third, to accurately measure $L_{X,GAS}$, we subtract the non-gas components with the two methods described in Section 4. Note that we cannot directly use $L_{X,GAS}$ to build the hot gas XLF because the source detection and the limiting sky area function were made in each energy band for all emission components included. Instead, based on the S-band XLF in Figure 9, we shift the data points to the left (lower $L_X$) after subtracting the non-gas components from $L_X(S)$ of the galaxies in each $L_X$ bin. We also shift the data points down (lower $\phi$) after excluding those with negative $L_{X,GAS}$. Note that the volumes applied in this step are correct because we use the same S-band for the sky area and source detection. However, to explore $L_{X,GAS}$ (S) instead of $L_X(S)$, we adjusted the data points and recalculated the errors with $L_{X,GAS}$ (S) and the number of galaxies in each bin.

In Figure 10, we show the results of Method A, subtracting $L_{X,LMXB}$ and $L_{X,HMXB}$ based on the scaling relations. The XLF in the entire redshift range ($z < 0.6$) is on the left panel, and the XLFs in three redshift bins are on the right panel. The hot gas XLFs shift slightly to the left and downward from the S-band XLFs but are almost identical except at the lowest $L_X$, which is unreliable. In Figure 11, we show the results of Method B, subtracting from $L_X(S)$ the power-law component estimated from the H-band. The hot gas XLF with Method B is further shifted to the left and downward because $L_{X,GAS}$ is further reduced, and those galaxies (~35%) with negative $L_{X,GAS}$ are excluded.

The general trends of the S-band XLF in Figure 9 remain valid in the hot gas XLFs made with the two subtraction methods. (1) The hot gas XLF ($z < 0.6$) is relatively well represented by a single power-law with a slope of -1 in the range $L_X(S) = 10^{40}$ and $10^{42}$ erg $s^{-1}$. (2) The number of galaxies (per unit co-moving volume) increases by a factor of a few from z = 0.1 to z = 0.3 in the range $L_X(S) = 10^{40} - 10^{41}$ erg $s^{-1}$. Similarly, the number of galaxies increases by a factor of a few from z = 0.3 to z = 0.5 in the range $L_X(S) = 10^{40.5} - 10^{41.5}$ erg $s^{-1}$.

At the high $L_X$ end, $L_X(S) > 10^{41.5}$ erg $s^{-1}$, no significant XLF change as a function of z is visible, although the error bars are large. We note that $L_X(S) = 10^{41.5}$ erg $s^{-1}$ corresponds to $L_X(B) > 10^{42}$ erg $s^{-1}$, higher than the luminosity of typical normal galaxies. We will discuss these luminous galaxies in Section 6.1.

To further quantify the z-dependent XLF changes, we fit the XLFs in $L_{X,GAS} = 10^{40} – 10^{42}$ erg $s^{-1}$ with single power-laws. The best-fit parameters and correlation coefficients are summarized in Table 2. As noted in the general trend of the XLF change, the best-fit power-law becomes steeper with increasing redshift. The significance of the difference between z=0.1 and z=0.5 is ~ 4-5$\sigma$ for both slope and intercept. In all cases, the correlation coefficient is close to -1.0, indicating a strong correlation. The last column in Table 2 is the value of $L_{X,GAS}$ on the best fit power-law at $\log \phi$ = -3.0 where the z-dependency can be reliably measured, i.e., in the range $L_{X,GAS} = 10^{40-41}$ erg $s^{-1}$. This indicates that $L_{X,GAS}$ decrease by a factor of 6-10 from z=0.5 to z=0.1.

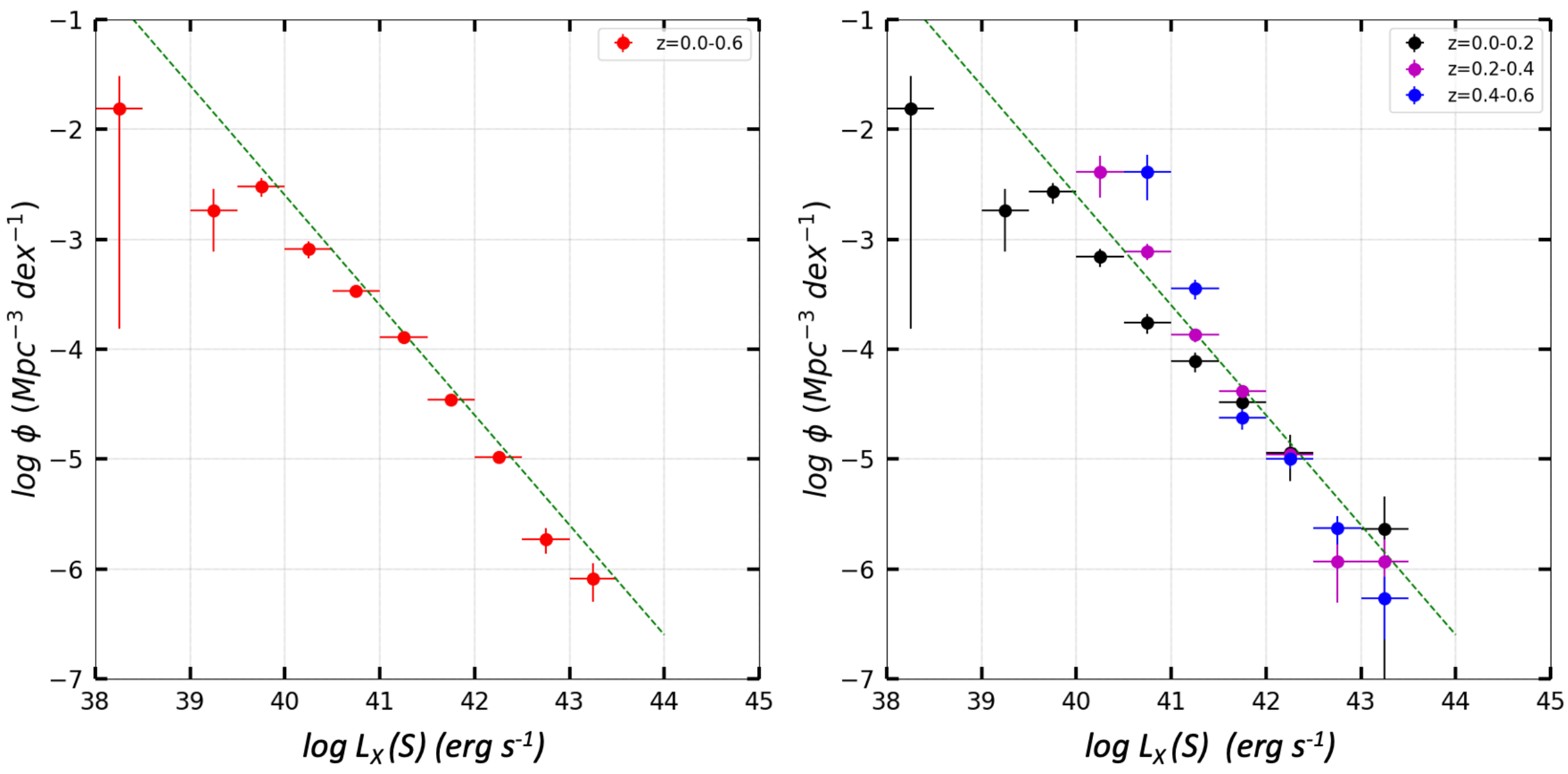

Figure 9. (left) The S-band (0.5-1.2 keV) XLF of galaxies with z < 0.6. (right) XLF separately built in three redshift bins: z=0.1 (0.0-0.2), z=0.3 (0.2-0.4), and z=0.5 (0.4-0.6).

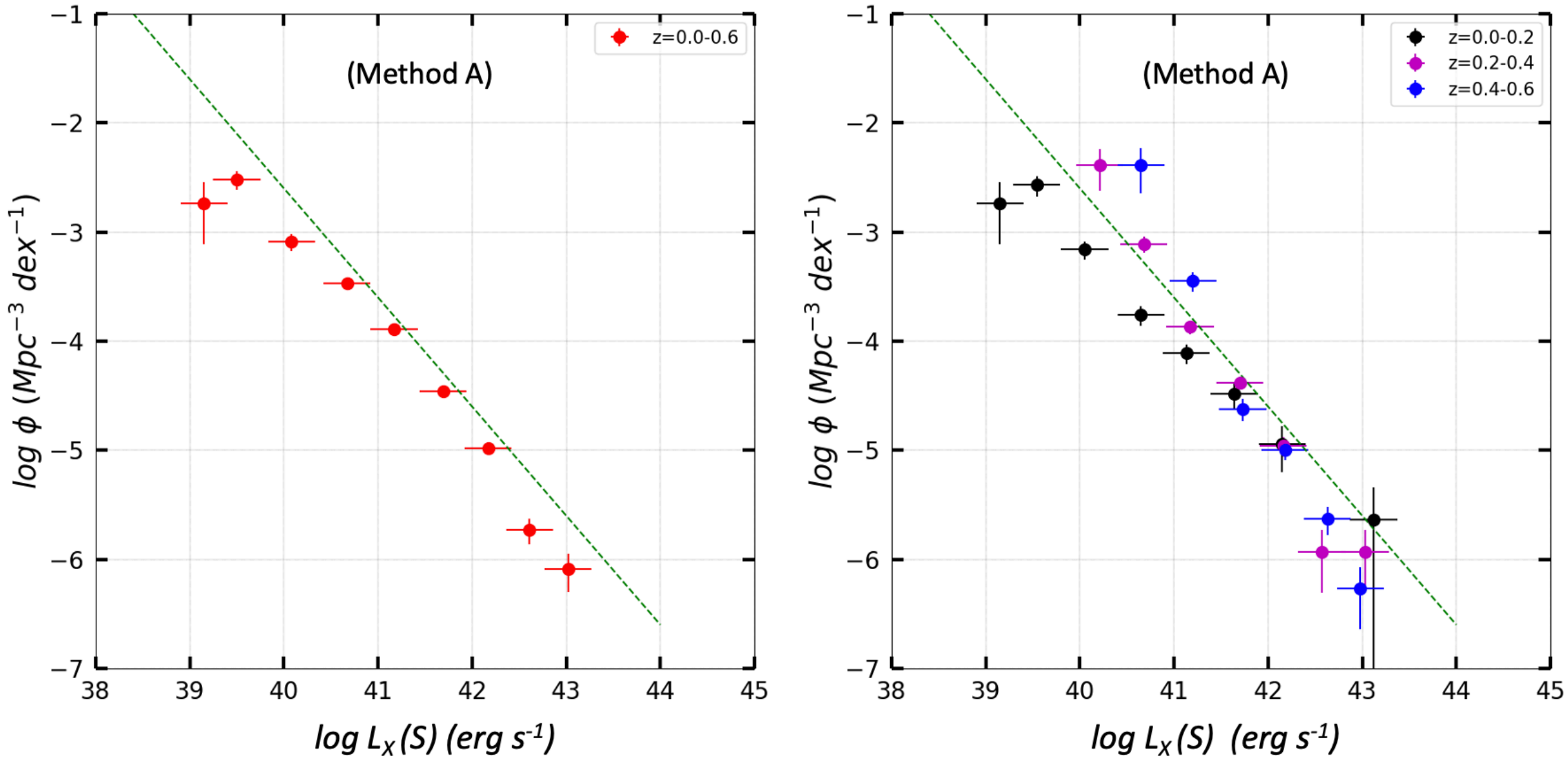

Figure 10. same as Figure 9, but with $L_{X,GAS}$ in the S-band determined with Method A in Section 4.1, subtracting $L_{X,LMXB}$ and $L_{X,HMXB}$ based on the scaling relations.

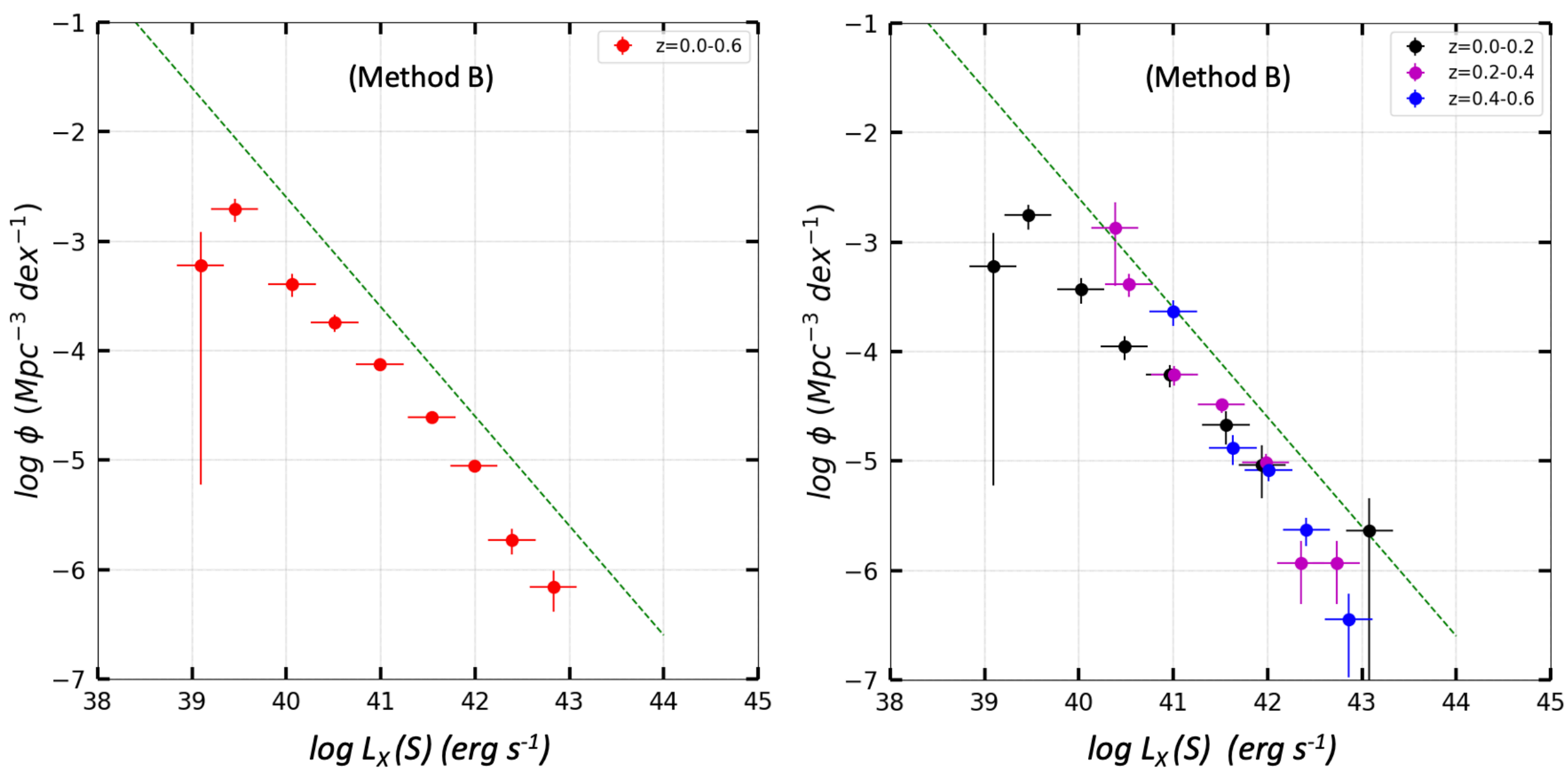

Figure 11. same as Figure 9, but with $L_{X,GAS}$ in the S-band determined with Method B in Section 4.2, subtracting the power-law component converted from the H-band to the S-band.

Table 2. Best-fit XLF parameters

| z | slope | intercept | rms | corr coefficient Pearson | Spearman | log $L_{X,GAS}$ at log $\phi$ =-3 |
|---|---|---|---|---|---|---|
| Method A | | | | | | |
| 0.0-0.6 | -0.97 (0.06) | 36.13 (2.57) | 0.08 | -1.00 | -1.00 | |
| 0.1 (0.0-0.2) | -0.91 (0.11) | 33.51 (4.62) | 0.07 | -1.00 | -1.00 | 40.12 |
| 0.3 (0.2-0.4) | -1.39 (0.04) | 53.48 (1.57) | 0.10 | -1.00 | -1.00 | 40.63 |
| 0.5 (0.4-0.6) | -1.68 (0.06) | 65.68 (2.81) | 0.18 | -0.98 | -1.00 | 40.87 |
| Method B | | | | | | |
| 0.0-0.6 | -1.02 (0.07) | 37.50 (2.78) | 0.12 | -0.99 | -1.00 | |
| 0.1 (0.0-0.2) | -0.91 (0.13) | 33.04 (5.31) | 0.10 | -0.99 | -1.00 | 39.60 |
| 0.3 (0.2-0.4) | -1.37 (0.13) | 52.39 (5.18) | 0.20 | -0.97 | -0.99 | 40.43 |
| 0.5 (0.4-0.6) | -1.87 (0.16) | 73.00 (6.79) | 0.22 | -0.97 | -1.00 | 40.64 |

## 6. DISCUSSION

### 6.1 The XLF of the hot gas component of galaxies with z < 0.6

In the $L_{X,GAS}$ range from $10^{40}$ erg $s^{-1}$ to $10^{42}$ erg $s^{-1}$, the overall hot gas XLF is well represented by a single power-law with best fit slope of -1.0 ± 0.1, no matter which method (Section 4) is used to estimate the hot gas component of the X-ray luminosity (see the left panels of Figures 10 and 11).

At the lower luminosities ($L_{X,GAS} < 10^{40}$ erg s$^{-1}$), the XLF is less accurate because only a small number of local (z < 0.2) galaxies are detected in the Chandra Source Catalog S-band (Figure 4) and because in these low-luminosity galaxies the contribution from the population of X-ray binaries is significant in the S-band (Figure 5). At the higher luminosities ($L_{X,GAS} > 10^{42}$ erg s$^{-1}$), the XLF is likely to be contaminated by obscured AGNs and galaxy groups (the 'XBONG'- X-ray bright optically normal galaxies, Kim et al. 2023b).

The presence of XBONGs is illustrated by Figure 12, where we plot the ratio of $L_X(S)$ to that of the power-law contribution $L_{X,PL}(S)$ [= $L_X$ (H)*C, Method B] against the broad band luminosity $L_X(B)$. $L_X(S)$ and $L_X(H)$ are K-corrected for hot gas and power-law models, respectively. $L_X(B)$ is not K-corrected because of the mixed emission components, but the K-correction will not significantly change the result. In Figure 12, 73% of the points with hard X-ray spectra (below the red horizontal line) have $L_X(B) > 10^{42}$ erg s$^{-1}$ (on the right side of the blue vertical line), indicating that they may be obscured AGNs. The points in the upper-right quadrant, with $L_X(S) > L_{X,PL}(S)$, and $L_X(B) > 10^{42}$ erg s$^{-1}$, may be AGNs optically diluted by the stellar light of massive host galaxies that, in the X-rays, are dominated by the emission of hot gaseous halos. Some of these objects could also be groups of galaxies, which would retain a larger amount of hot gas than normal galaxies (see Kim et al. 2023b). There are 81 objects with soft X-ray spectra which are not shown in Figure 12 because they are not detected in the H-band. 32 of them have $L_X(B) > 10^{42}$ erg s$^{-1}$ and they are also likely to be large halo or group galaxies.

Because of these effects, below we will only discuss the $L_{X,GAS}$ range from $10^{40}$ erg s$^{-1}$ to $10^{42}$ erg s$^{-1}$.

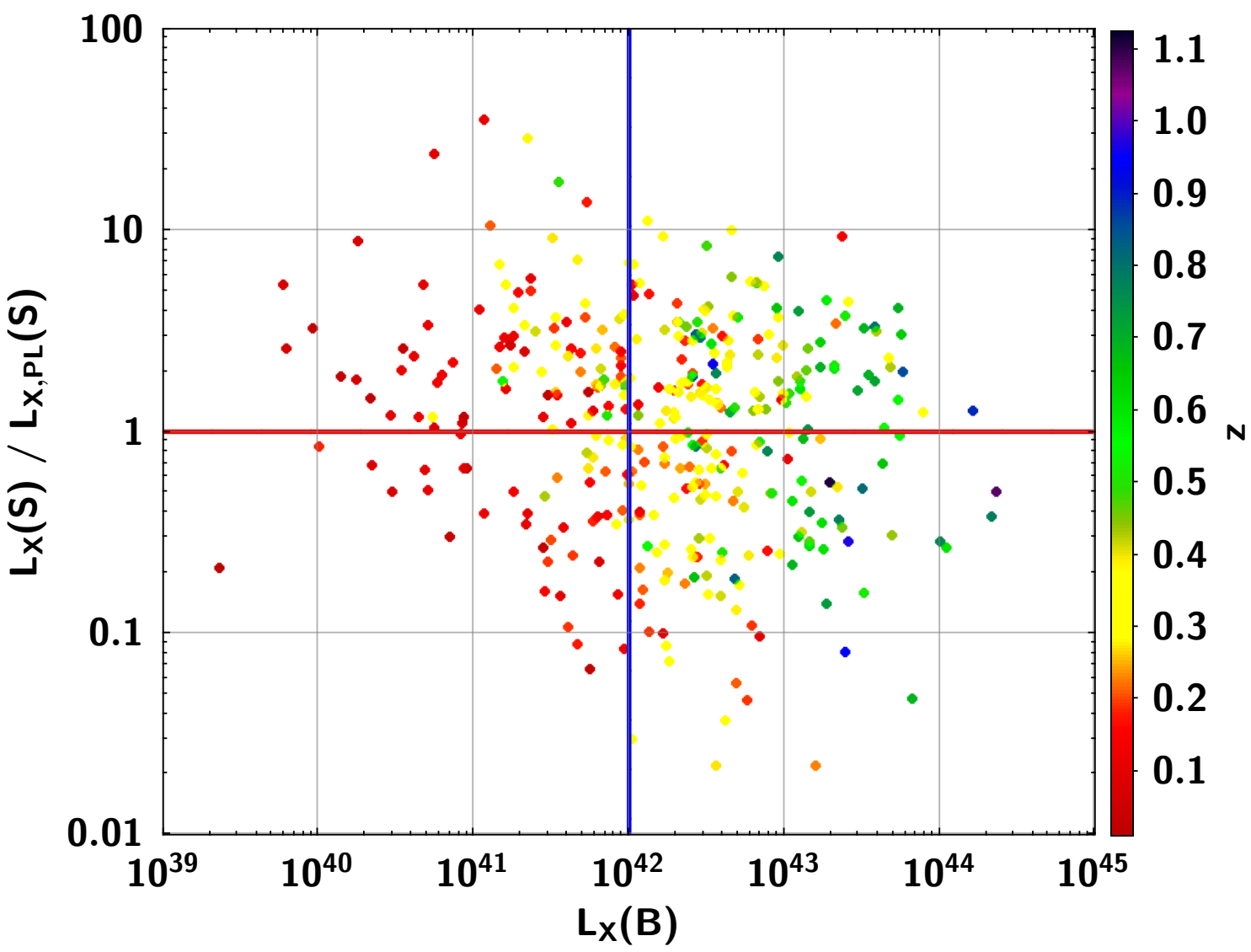

Figure 12. The ratio of $L_X(S)$ and $L_{X,PL}(S)$ plotted against $L_X(B)$. The data points are color-coded by z. The horizontal red line indicates ratio = 1, and the blue vertical line indicates $L_X(B) = 10^{42}$ erg s$^{-1}$, which marks the canonical boundary between normal galaxies and AGNs.

6.2 The redshift evolution of the $10^7$ K hot gas component of normal galaxies with $L_{X,GAS} \sim 10^{41}$ erg $s^{-1}$

Table 2 (Section 5) summarizes the results of the power-law fits of the hot gas XLFs derived for three different redshift bins from z = 0.0 to 0.6. The XLF becomes progressively steeper at higher z (slope ≃ -1.4 and -1.8 at z = 0.3 and 0.5, respectively), suggesting a larger number of lower-luminosity galaxies (per unit volume) at higher redshifts. The steepening of the XLF of the hot gas luminosity $L_{X,GAS}$ with z suggests that the time scale of the luminosity evolution depends on $L_{X,GAS}$ in the sense that the decrease in the amount of hot gas with decreasing z may be more severe in the lower $L_{X,GAS}$ galaxies.
Comparing the local universe XLF to those in higher redshift bins (Figure 11), we find a 5σ increase in the number of galaxies with $L_{X,GAS} \sim 10^{40.5\text{-}41.0}$ erg $s^{-1}$ with increasing redshift. If the change in the XLF is due to pure luminosity evolution, $L_{X,GAS}$ decreases by a factor of 6-10 for the last 5 Gyr (from z = 0.5 to 0.1).

In typical cooling flow models of early-type galaxies without AGN feedback, the hot gas luminosity $L_{X,GAS}$ gradually decreases with time after rapid changes caused by early type II SN driven galactic winds (e.g., Loewenstein & Mathews 1987; David et al. 1991; Ciotti et al. 1991), primarily because of the continuous decrease of the stellar mass loss rate (Pellegrini 2012). Based on the early cooling flow models, $L_{X,GAS}$ typically changes about a factor of 2-3 in the last 5 Gyr. Our result is qualitatively consistent with this predicted change, but the amount of change implied by the z-dependence of the hot gas XLF is more than predicted. Recent simulations with AGN feedback produced a similar overall trend, decreasing $L_{X,GAS}$ with time, when the nucleus is inactive. However, sporadic AGN outbursts cause $L_{X,GAS}$ to fluctuate significantly (e.g., Choi et al. 2014, Ciotti et al. 2017). In this case, the duty cycle and duration of AGN outbursts may be more important than the gradual change of the mass loss rate to determine the $L_{X,GAS}$ evolution.

Mineo et al. (2012) examined the Chandra data of nearby late-type galaxies and demonstrated that in this sample the X-ray luminosity of the hot gas is linearly correlated with the star formation rate. Taking their relation and the cosmic star formation history which peaks at z=2-3 (e.g., Madau & Dickson 2014), we expect the $L_{X,GAS}$ in late-type galaxies to increase with increasing z. Quantitatively, because the star formation rate in a given co-moving volume decreases by a factor of ~3 for the last 5 Gyr, we expect the $L_{X,GAS}$ decreases by a factor of ~3 for the last 5 Gyr.

Our result of the $L_{X,GAS}$ change with redshift based on the XLF study is qualitatively consistent with the expected change for both early-type and late-type galaxies, but the amount of change implied by the z-dependence of the hot gas XLF is larger than expected.

The steepening of the $L_{X,GAS}$ XLF with increasing z may have implications for the missing baryon problem in galaxies (e.g., Tumlinson, et al. 2017). An XLF slope of -1, as observed in the full sample (left panel of Figures 10 and 11) indicates that the total amount of hot gas is evenly distributed between less luminous (less massive) and more luminous (more massive) galaxies. The flatter slope (-0.9) at z = 0 suggests that the hot gas preferentially resides in larger galaxies. Because of the steeper XLF slope, the opposite is true at higher z. With the best-fit XLF slope of -1.4 (-1.8) at z = 0.3 (0.5), the number of galaxies at $L_{X,GAS} = 10^{40}$ erg $s^{-1}$ is about 25 (63) times higher than that at $L_{X,GAS} = 10^{41}$ erg $s^{-1}$. In other words, more hot gas resides in lower $L_{X,GAS}$ galaxies at higher z. This result could in part explain the difficulty of finding enough hot gas at z = 0 and emphasizes the importance of higher z, moderate $L_{X,GAS}$ galaxy samples to search for the missing baryons.

### 6.3 The more efficient gas retention in larger galaxies

As discussed above, the flatter XLF slope ($\simeq$ - 0.9) at z = 0.1 suggests a relatively small local excess of high luminosity galaxies. Moreover, the XLFs at different z in the range of $L_{X,GAS} = 10^{41.5 - 42}$ erg $s^{-2}$ are consistent with each other, suggesting either lack of luminosity evolution or a small increase with time of the hot gas luminosity.

Because $L_{X,GAS}$ is closely related to the total mass (including dark matter) of the galaxy, $M_{TOTAL}$ (Kim & Fabbiano 2013), this result may be explained by the effect of deeper potential wells on the retention of the hot gas component. This conclusion is consistent with the lack of XLF evolution found in the massive galaxy groups and clusters, dominated by even larger potential wells, to z ~ 0.6 (e.g., Boehringer et al. 2002; Finoguenov et al. 2020).

### 6.4 The $L_{X,GAS}$ - $M_{STAR}$ Relation

Unlike X-ray binaries (Section 4), the hot gas luminosity is not directly related to the stellar mass. Although most likely originating from stars, the amount of hot gas retained by a galaxy can be significantly affected by other factors such as the total mass (including dark matter), AGN/stellar feedback, and environmental effects (Kim & Fabbiano 2013; Nardini, Kim & Pellegrini 2022). Since the first X-ray imaging mission - the Einstein X-ray Observatory - detected the X-ray emission of galaxies, it has been known that there is a large scatter in the relation between the total X-ray ($L_{X,TOT}$) and optical luminosities of both early-type and late-type galaxies (e.g., Fabbiano, Kim & Trinchieri 1992). When $L_{X,GAS}$ was accurately measured after excluding X-ray binaries with Chandra observations, the scatter in the $L_{X,GAS}$ - $L_K$ relation (where $L_K$ is the luminosity in the K band) was found to be even larger than that in the $L_{X,TOT}$ - $L_K$ relation, reaching a factor of ~1000 for a given optical luminosity or stellar mass (e.g., Kim & Fabbiano 2013, 2015).

With the large sample assembled in this work, we revisit this relation, keeping in mind that the present sample includes both early- and late-type galaxies. In the left panel of Figure 13, we plot the total soft band $L_X(S)$ against $M_{STAR}$. The K-correction for the APEC model is applied. Each data point is color-coded by z. The scatter in $L_X$ for a given $M_{STAR}$ is indeed large, and the possible correlation, if any, is very steep. The correlation coefficients from Pearson and Spearman tests are 0.29 and 0.28, respectively, indicating a weak or negligible correlation. Given that only high $L_X$ galaxies are detected at high z, we run the tests again with a sub-sample of local galaxies at z < 0.2 (most red points in Figure 13), where both optical and X-ray observations are near-complete. The correlation coefficients are 0.19 and 0.21, indicating an even poorer correlation.

The diagonal lines in Figure 13 indicate the expected $L_{X,LMXB}$ (solid) and $L_{X,HMXB}$ (dashed) contributions at z=0 (red) and at z=0.6 (green). The red data points (z < 0.2) can be compared with the red lines (z = 0), which roughly outline the bottom of the $L_X(S)$-$M_{STAR}$ distribution. The green data points (z ~ 0.6) instead, are well above the green lines (z=0.6), given the limiting flux of the Chandra observations.

In the right panel of Figure 13, we also plot the total hard band $L_X$(H) against $M_{STAR}$ to show the expected contribution from the X-ray binaries in the H-band. The K correction for the PL model is applied. The diagonal lines are the same as in the left panel but scaled up for the different energy bands. Again, the red lines are close to the lower luminosity end in $L_X$(H) = $10^{39}$ - $10^{40}$ erg $s^{-1}$ for those galaxies where the X-ray binaries dominate the H-band. Considering 81 galaxies (of 453) which are not detected in the H-band, about one-quarter may have $L_X$(H) consistent with expected $L_{X,LMXB}$(H) + $L_{X,HMXB}$(H). The galaxies with excess $L_X$(H) (> ~5 x $10^{41}$ erg $s^{-1}$) are likely obscured AGNs, as described in Section 6.1 and Figure 12. This justifies the necessity of Method B.

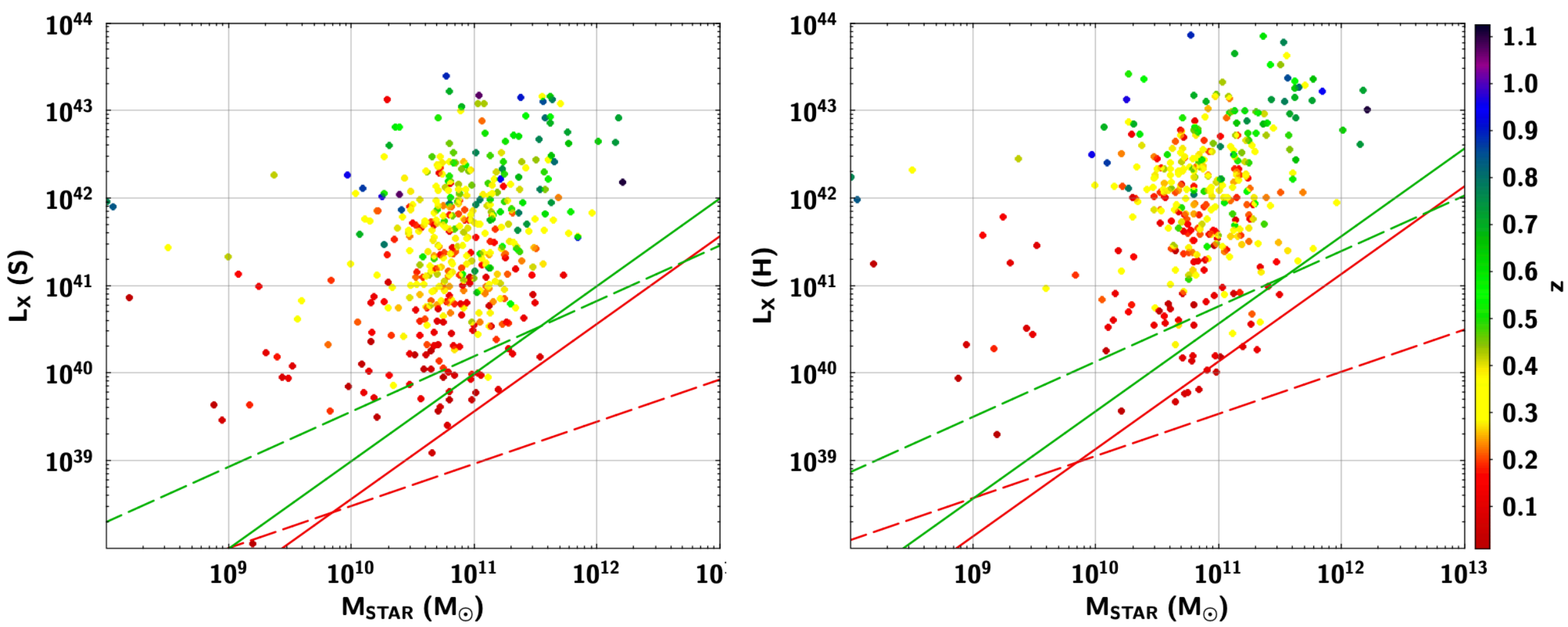

Figure 13. $L_X$ in the S-band (left) and in the H-band (right) plotted against $M_{STAR}$. The K-correction for the APEC (left) and PL model (right) is applied. Each data point is color-coded by z (see color scale on the right side of the figure). The red solid and dashed lines indicate the expected $L_{X,LMXB}$ and $L_{X,HMXB}$ at z=0, respectively. The green solid and dashed lines indicate the expected $L_{X,LMXB}$ and $L_{X,HMXB}$ at z=0.6, respectively.

In Figure 14, we plot $L_{X,GAS}$, determined by the two methods in Section 4, against $M_{STAR}$. The scatter in $L_{X,GAS}$ for a given $M_{STAR}$ is even larger than when using the total soft band luminosity $L_X$(S). This is primarily because galaxies with low $L_X$(S) tend to have a relatively higher contribution of X-ray binary luminosity, while the emission of galaxies with high $L_X$(S) is dominated by their hot halos, and the applied correction is minimal. The correlation coefficients from the Pearson and Spearman tests are 0.28 and 0.25, respectively. For the local sub-sample (z < 0.2), the correlation coefficients are 0.24 and 0.23. We note that the selection bias affects the scaling relation because we do not consider the X-ray undetected galaxies. The stacking analysis, described below, shows a roughly comparable $L_X$(S) for a given $M_{STAR}$, indicating that the selection bias is small.

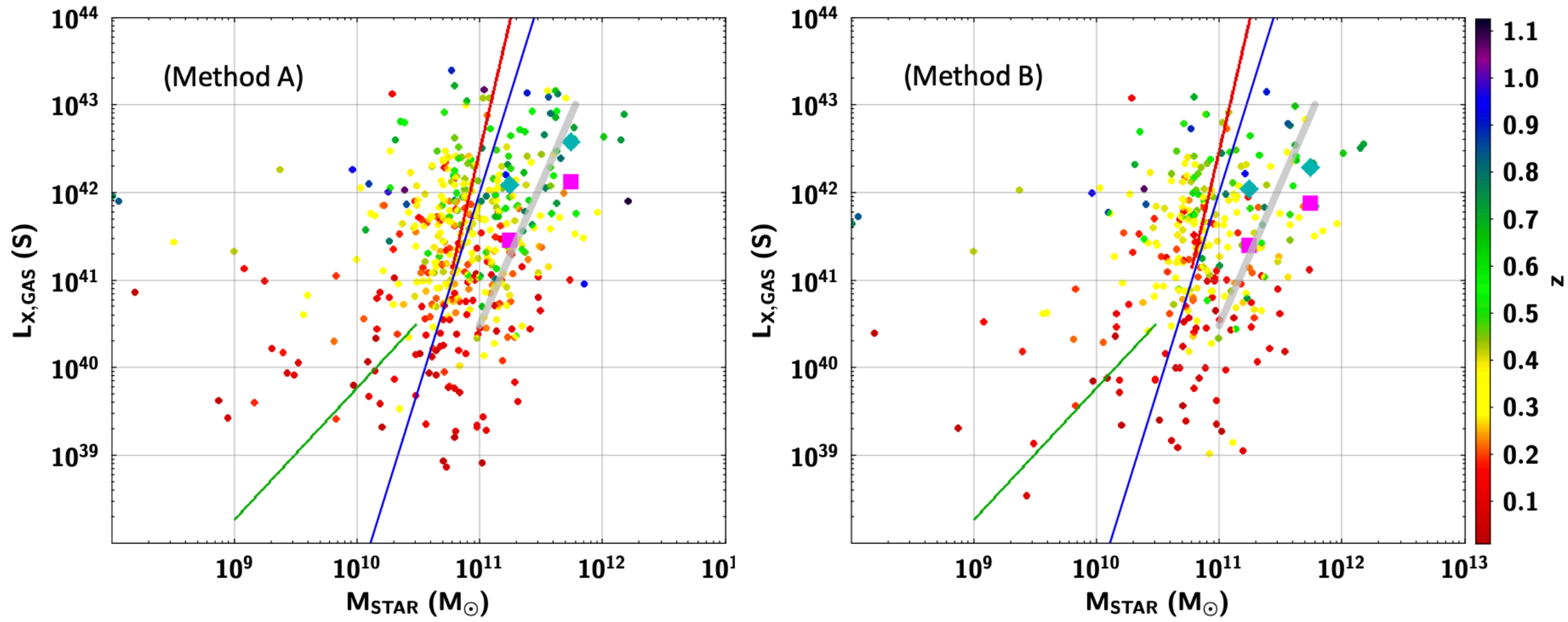

Figure 14. same as Figure 13 (left), but with $L_{X,GAS}$ determined with Method A (left) and Method B (right) in Section 4. The two steep lines indicate $L_{X,GAS} \sim M_{STAR}^{6}$ (red) and $L_{X,GAS} \sim M_{STAR}^{4.5}$ (blue). The flatter green line indicates $L_{X,GAS} \sim M_{STAR}^{1.5}$. The thick transparent grey line indicates the stacking result from Anderson et al. (2015). The green and magenta points represent average values in the $M_{STAR}$ range used by Anderson et al. See the text for the details.

Using local early-type galaxies, Kim & Fabbiano (2013) found a steep relation between $L_{X,GAS}$ and $L_K$ (or $M_{STAR}$), $L_{X,GAS} \sim L_K^{4.5}$, but with a large scatter. See Li & Wang (2013) for a large scatter in local late-type galaxies and Jones et al. (2014) for the scatter plot of $L_X$(full) and $M_{STAR}$ for multiple types of galaxies (based on the 4000A break). With early-type galaxies in the COSMOS survey (Elvis et al. 2009), Civano et al. (2014) suggested a similarly steep relationship. This relation corresponds to $L_{X,GAS} \sim M_{STAR}^{4.5}$, assuming $M_{STAR} \sim L_K$ in early-type galaxies (Bell et al. 2003). The blue line in Figure 14 indicates this relation with an arbitrary normalization. Attempts to find a possible correlation between $L_X$ and $M_{STAR}$ were also made by stacking X-ray data for large samples of optically selected galaxies. Paggi et al. (2016) stacked the Chandra data of the early-type galaxies in the COSMOS field and found a similar result as in the sample of X-ray detected galaxies in the COSMOS field (Civano et al. 2014) after excluding $L_{X,LMXB}$. They also found that X-ray excess ETGs above the relation often exhibit hard spectra, indicating the presence of hidden, absorbed AGNs at the high $L_X$ end. Anderson et al. (2015) stacked the RASS (ROSAT all-sky survey) data of SDSS locally bright galaxies at $z < 0.4$ and found $L_{X,TOT}$ (0.5-2 keV) $\sim M_{STAR}^{3.3}$ at the high mass end, $M_{STAR} = 10^{11}$ - $10^{12}$ $M_{\odot}$ where $L_{X,LMXB}$ and $L_{X,HMXB}$ are not significant. Comparat et al. (2022) stacked the eRosita eFEDS data of the GAMA galaxies at $z < 0.3$ and found a similar relation as Anderson et al. (2015) at the high mass end, $M_{STAR} > 10^{11}$ $M_{\odot}$, indicated by the thick transparent grey line in Figure 14. While this relation has a power-law trend compatible with the scaling relations derived from the scatter diagram, it hides the large scatter we find when plotting individual detections. It is consistent with the mean $L_{X,GAS}$ we derive from the scatter diagram in the same range of $M_{STAR}$, represented by the cyan diamonds (arithmetic mean) and magenta squares (geometric mean).

The $L_{X,GAS}$ - $M_{STAR}$ noisy relation could be a byproduct of the much tighter correlation between $L_{X,GAS}$ and $M_{TOT}$ (= $M_{STAR}$ + $M_{DM}$), found by Kim & Fabbiano (2013), where $M_{TOT}$ was measured from the GC kinematics of individual galaxies. This tight relation, $L_{X,GAS} \sim M_{TOT}^3$ for early-type galaxies with a small rms of ~0.1 dex, suggests that the total mass is the primary factor in regulating the amount of hot gas (see also Forbes et al. 2017; Kim et al. 2019). We also note that in late-type galaxies $L_{X,GAS}$ is tightly correlated with SFR rather than $M_{STAR}$ (e.g., Mineo et al. 2012). Assuming the stellar-halo mass relation (e.g., Behroozi et al. 2013), which can be represented by a broken power law, from the $L_{X,GAS} \sim M_{TOT}^3$ we can approximately estimate the expected relation between $L_{X,GAS}$, and $M_{STAR}$. At the high mass end ($M_{STAR} > 3 \times 10^{10}$ $M_\odot$), $M_{HALO} \sim M_{STAR}^2$ makes the relation very steep, $L_{X,GAS} \sim M_{STAR}^6$, as indicated by the red line in Figure 14 (with an arbitrary normalization). At the low mass end ($M_{STAR} < 3 \times 10^{10}$ $M_\odot$), $M_{HALO} \sim M_{STAR}^{0.5}$ results in a considerably flatter relation, $L_{X,GAS} \sim M_{STAR}^{1.5}$, indicated by the green line in Figure 14. While this broken-line approximation follows the observed data, the scatter is large. As Figure 14 shows, $L_{X,GAS}$ varies from $10^{39}$ to $10^{42}$ erg $s^{-1}$ among local ($z < 0.2$) galaxies with a range of $M_{STAR}$ = a few x $10^{10}$ - $10^{11}$ $M_\odot$. $M_{STAR}$ cannot be used for determining the gas content and luminosity $L_{X,GAS}$ of any given galaxy, although there is a general trend of larger $L_{X,GAS}$ being associated with larger total mass galaxies that also have a larger stellar mass.

## 7. SUMMARY AND CONCLUSIONS

The main points of this paper are summarized below:

- This study makes use of 653 galaxies in the Chandra Galaxy Catalog (Kim et al. 2023a), extracted from five Chandra large-area surveys inside the SDSS footprint. These areas are outside the Galactic plane and away from any known complex objects, hence optimal for the XLF study of normal galaxies. This is the largest sample so far used for systematic studies of the X-ray properties of galaxies.
- Using the 575 galaxies with $z < 0.6$, we built a broad B-band (0.5-7.0 keV) X-ray luminosity function (XLF) and compared it with those in the literature from much small samples of galaxies, finding good agreement.
- We then estimated the X-ray luminosity of the hot gas component of our sample galaxies, $L_{X,GAS}$ by subtracting the contribution of the X-ray binary population and possible AGN from the 0.5-1.2 keV S-band luminosity of the Chandra Source Catalog CSC2.0 (Evans et al. 2010), $L_X(S)$.
- We followed two methods to estimate the non-gaseous X-ray emission in the S-band: (A) We used the scaling relations found for LMXB and HMXB populations in Chandra studies (Lehmer et al. 2016); (B) We assumed that the 2.0-7.0 keV CSC2.0 H-band luminosity $L_X(H)$ is entirely dominated by the X-ray binary and AGN emission, which we approximated as a power-law spectral component, from which we estimated its contribution in the S-band. The two estimates differ within a factor of ~3, and effectively provide the upper and lower boundaries of $L_{X,GAS.}$ X-ray luminous galaxies with $L_X$(0.5-7 keV) > $10^{42}$ erg $s^{-1}$ tend to give negative $L_{X,GAS}$ values when estimated with method B. These galaxies are part of the XBONGs class (Kim et al 2023b) of likely obscured AGNs.
- We built the XLF of $L_{X,GAS}$ in three redshift bins ($z$ = 0.1, 0.3, and 0.5) and found that the number of galaxies per unit co-moving volume increases ($5\sigma$) with increasing redshift, in the range of $L_{X,GAS}$ $10^{40}$ to $10^{41.5}$. No difference among the three z bins for $L_{X,GAS} > 10^{41.5}$ erg $s^{-1}$ is observed, within the errors. This steepening of the XLF at higher z is confirmed by fitting with a single power-law in the range of $L_{X,GAS} = 10^{40} – 10^{42}$ erg $s^{-1}$ (which excludes XBONGs).

- The flatter slope (-0.9) at z = 0 suggests that the hot gas preferentially resides in larger galaxies. Because of the steeper XLF slope, the opposite is true at higher z. This XLF steepening suggests that there are relatively more $L_{X,GAS}$~$10^{41}$ erg $s^{-1}$ galaxies at higher z, suggesting that high z, moderate $L_{X,GAS}$ galaxies are the optimal target to solve the missing baryon problem.
- The redshift evolution of $L_{X,GAS}$ suggested by our study is qualitatively consistent with the expected change for both early-type galaxies, caused by the continuous decrease of the stellar mass loss rate (Pellegrini 2012), and late-type galaxies, caused by the z-dependence of the SFR (Madau & Dickson 2014). However, the change implied by the z-dependence of the hot gas XLF is a few times larger than expected.
- Our results confirm the huge scatter in the $L_{X,GAS}$ - $M_{STAR}$ relation and indicate that the integrated stellar mass of a galaxy cannot be used to determine the hot gas content, while there is a general trend of larger $L_{X,GAS}$ being associated with larger mass galaxies that also have a larger stellar mass.

## Appendix A. XLFs for red and blue galaxies

We attempted to explore the XLF separately in early-type and late-type galaxies. We tried multiple parameters to separate ET and LT galaxies, including optical spectral class and subclass, optical colors, WISE IR colors, PanSTARRS type, Legacy type (based on radial profile), and galaxy ZOO classifications. We found that different parameters indicate different results, e.g., optical spectroscopic (emission lines), photometric (colors), and morphological (ZOO) indicators do not point to a consistent type. Even if SDSS spectra exist, the fraction of galaxies with class=GALAXY and subclass(es) indicating star-formation significantly decreases with increasing z. The fraction of SF galaxies at z=0.5 is a factor of a few less than that at z=0.1. Similarly, the Galaxy Zoo (either by citizen scientists or AI) suffers a similar problem: the fraction of ETGs (i.e., no feature) increases with increasing z. In a following paper, we plan to explore how to reliably separate ET and LT by applying, with a solid statistical base, the Machine Learning Technique (e.g., Random Forrest Classifier) with multi-wavelength data.

Nonetheless, in this appendix we present the XLFs made separately for the red and blue galaxies, which may represent, to first-order approximation, quiescent and star-forming galaxies, respectively. We separate our samples into two groups, red and blue galaxies, by setting the boundary at g-r = 0.7, where g and r are K-corrected rest frame magnitudes. We show the g-r distribution of our sample in Figure A-1. This choice roughly corresponds to the division of galaxy color bimodality (e.g., Baldry et al. 2004; Nelson et al. 2018) and results in two subsamples with similar numbers of galaxies.

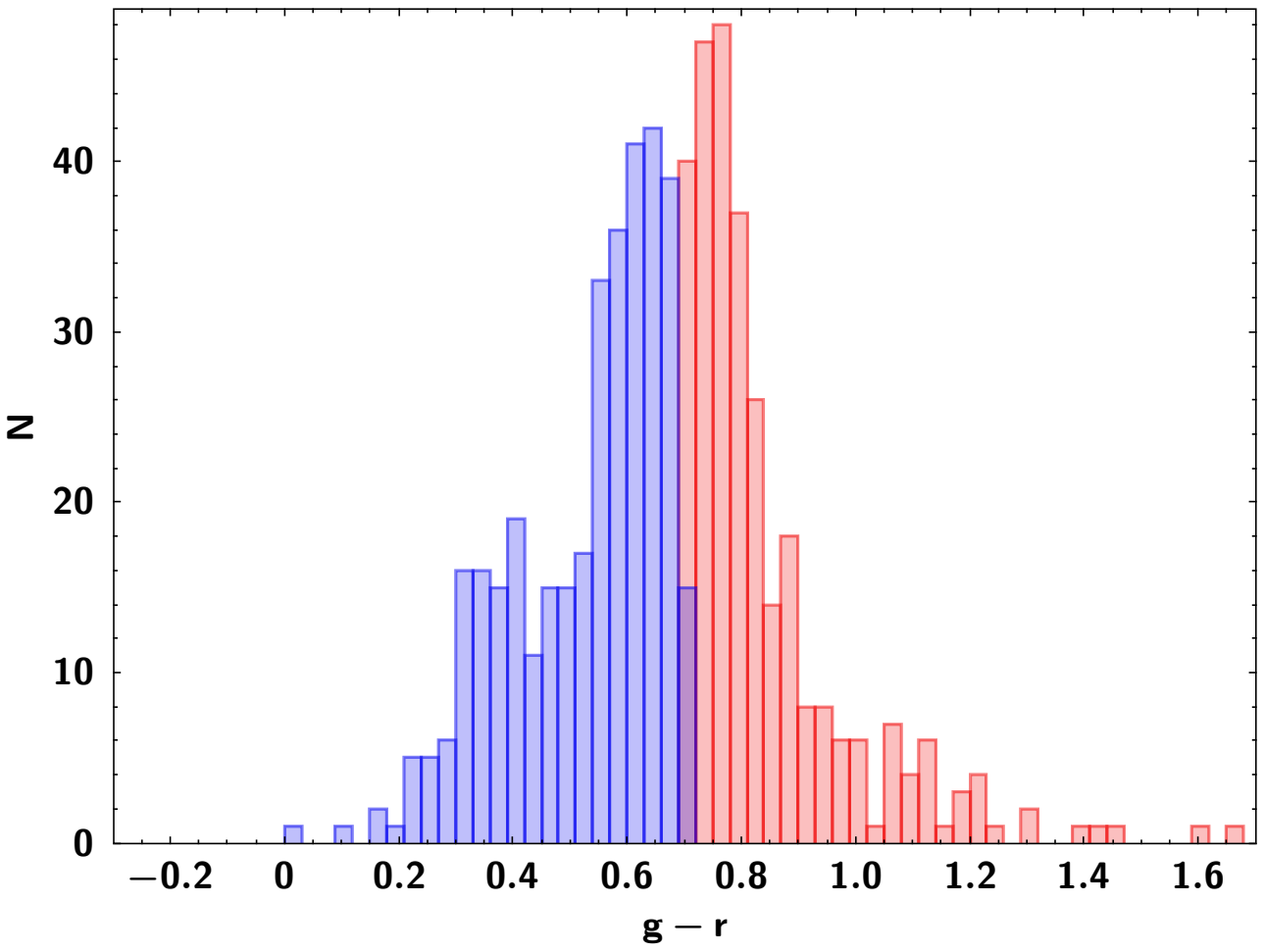

Figure A-1. g-r histogram of our sample. The boundary of red and blue subsamples is set at g-r = 0.7.

Figures A-2 and A-3 show the XLFs of the red and blue galaxies, respectively. With Method A, the XLFs of red and blue galaxies are similar to each other within the errors. With Method B, the XLFs of red and blue galaxies are subject to larger errors even at $L_{X,GAS} > 10^{40}$ erg $s^{-1}$, because of the smaller number of objects in each sub-sample. Again, we do not find any significant difference within the errors between the two XLFs. In both samples, we confirm that the number of galaxies in the $L_{X,GAS}$ range of $10^{40\text{-}41.5}$ erg $s^{-1}$ increases with z, although the two high z bins are not distinguishable in Method B.

We have also applied a u-r cut instead of g-r to cover wider wavelengths and more effectively pick star formation at the u-band. The results do not change significantly.

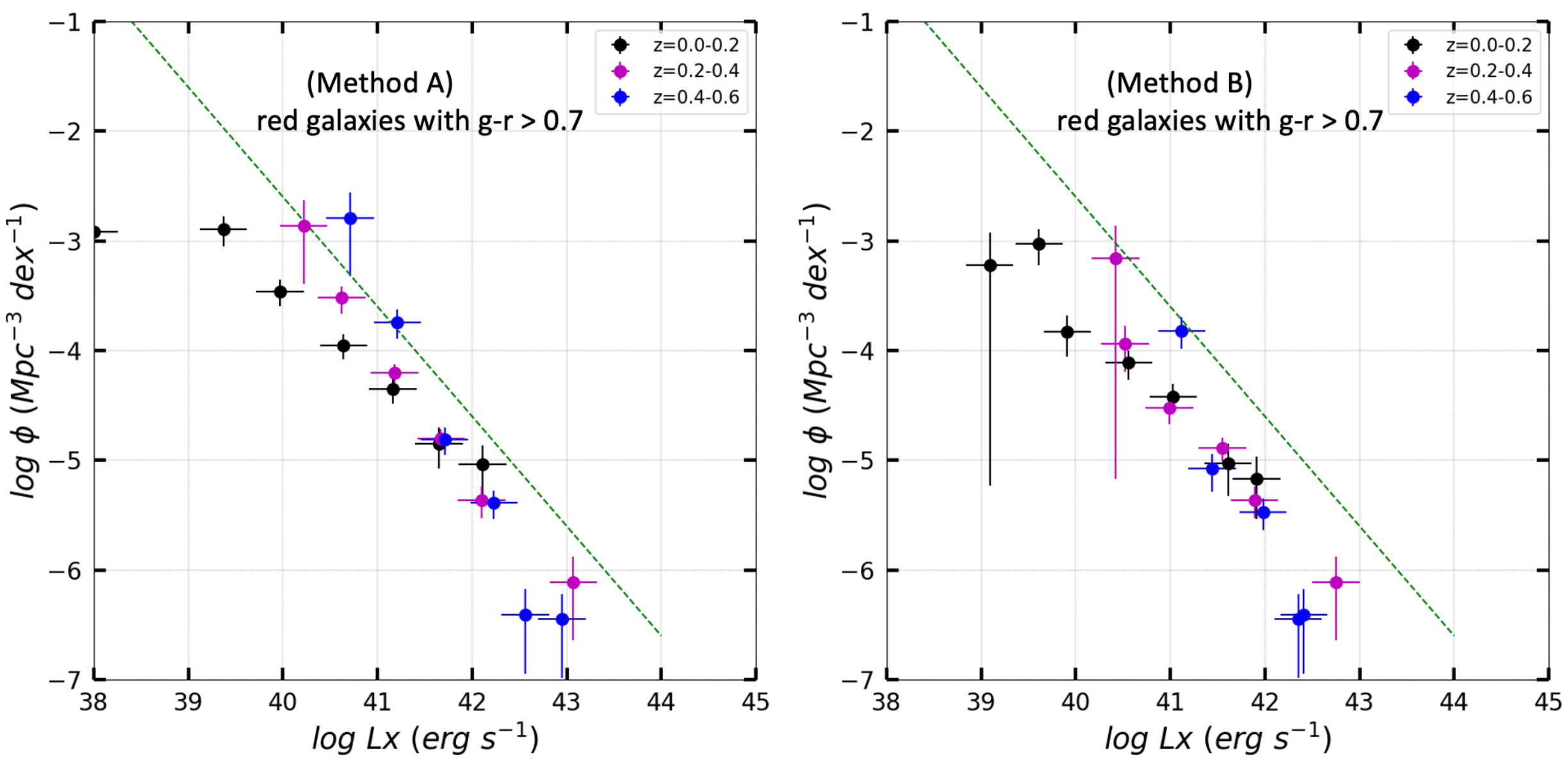

Figure A-2. same as the right panel of Figures 10 and 11, but only with red galaxies (g-r > 0.7).

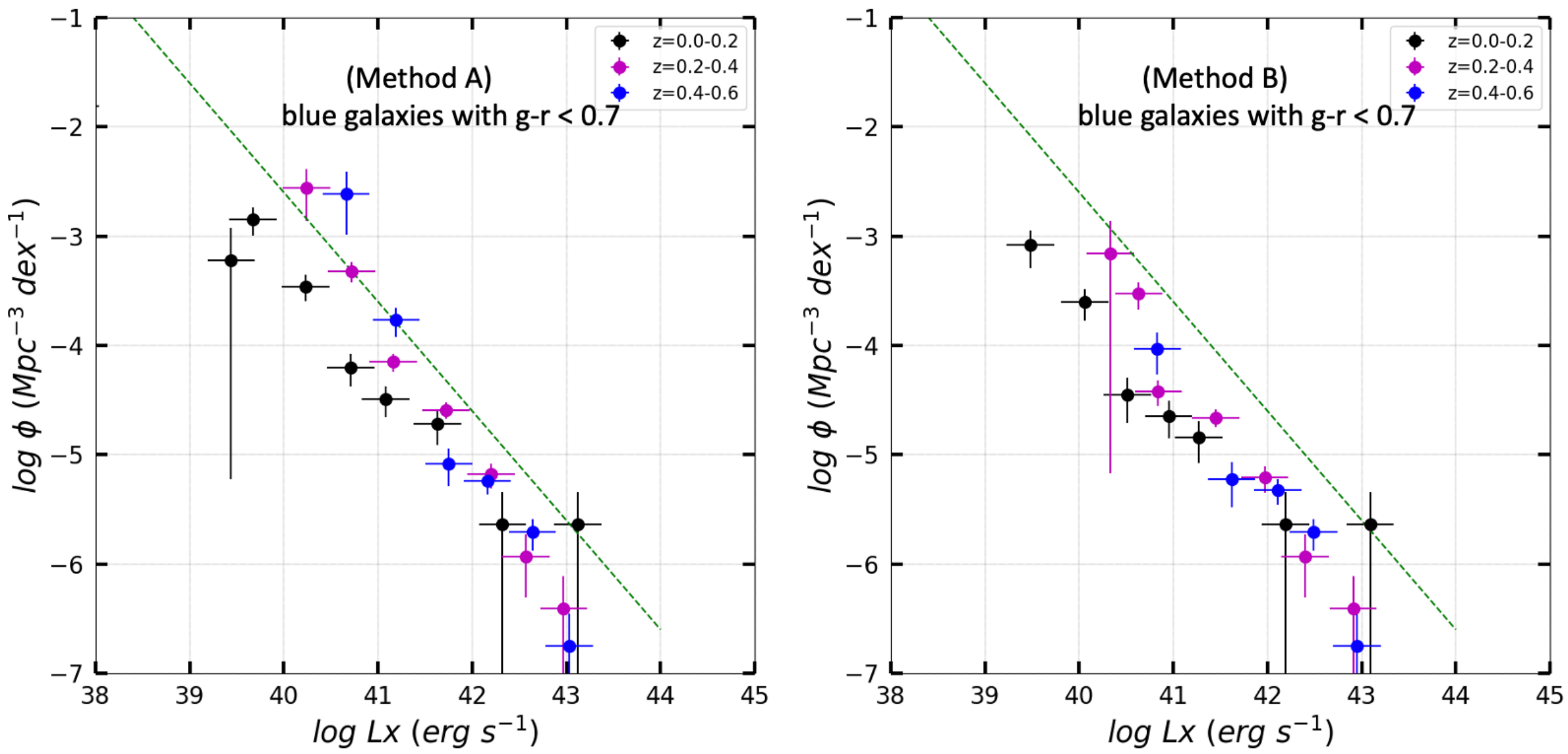

Figure A-3. same as Figure A-2, but only with blue galaxies (g-r < 0.7).

## Appendix B. Uncertainties in $L_{X,GAS}$

Here we consider the uncertainties in the estimate of $L_{X,GAS}$ (S) and their impacts on the XLFs. They include errors in (1) photometric redshift, (2) $L_{X,LMXB}$ (S), (3) $L_{X,HMXB}$ (S), (4) $T_{GAS}$, (5) $L_{X,PL}$ (S), and (6) crossing-over the $L_X$ bins after subtracting non-gas components.

To assess the uncertainty in photometric redshifts, Kim et al. (2023a) compared them with spectroscopic redshifts. All the available quality flags (e.g., zWarning, photoErrorClass, extrapolation_Class – for details, we refer to Kim et al. 2023a) were applied in this work. Kim et al. (2023a) removed low-quality data with error(z)/z < 0.5 from the sample. The outlier fraction of 5% was determined within 0.2 dex by comparing with spec-z. The outliers are mostly found at lowest and highest redshifts, which are beyond our z range of interest. Photo-z follows spec-z well in the range z=0.03-0.6 (see Figure C1 in Kim et al. 2023a), and the mean and standard deviation of the redshift ratio is $1.0 \pm 0.23$, assuming a normal distribution. The $1\sigma$ error of photo-z is 0.1 dex, which is smaller than the z bin size in the present work, and the corresponding error in luminosity (0.2 dex) is also smaller than the $L_X$ bin size (0.5 dex). Therefore, the photo-z uncertainty does not seriously affect our result.

The uncertainty in $L_{X,LMXB}$ (S) is minimal, given the rather tight correction between $L_{X,LMXB}$ and $M_{STAR}$. The $1\sigma$ rms scatter of this relation is 0.1 dex (Boroson, et al. 2011). As shown in Figure 5 (left), the $L_{X,LMXB}$ contribution will only affect galaxies with lower $L_{X,GAS}$ ($< 10^{40}$ erg $s^{-1}$).

The uncertainty in $L_{X,HMXB}$ (S) is primarily from the error in the SFR. The relation between SFR and $M_{STAR}$ in the galaxy main sequence has a typical $1\sigma$ width of $\pm 0.2$ dex (Speagle et al. 2014). Since $L_{X,HMXB}$ is proportional to SFR, the error in $L_{X,HMXB}$ is 0.2 dex. As shown in Figure 5 (right), this uncertainty only affects galaxies with lower $L_{X,GAS}$ ($\sim 10^{39}$ erg $s^{-1}$). For early-type galaxies where the galaxy main sequence is not applicable, subtracting $L_{X,HMXB}$ (S) is unnecessary. As in Figure 5 (right), it only affects a few high-mass but low-$L_X(S)$ galaxies, i.e., those points on the right side just above the red line. But those are at $z > 0.6$ and are not included in the XLFs.

We estimate $T_{GAS}$ based on the relation between $T_{GAS}$ and $L_{X,GAS}$ relation. While this relation is tight in ETGs (Kim & Fabbiano 2015), it is less tight in the LTGs (e.g., Li & Wang 2013). $T_{GAS}$ may vary between 0.4 and 0.7 keV for late-type galaxies with $L_{X,GAS} = 10^{40}$ erg $s^{-1}$. The temperature error only affects the K-correction in $L_X(S)$. In Figure 3, the error is ~0.1 dex at $z = 0.5$ and is almost negligible at $z < 0.4$.

In summary. the combined error from the above effects is < 0.3 dex for $L_{X,GAS} > 10^{40}$ erg $s^{-1}$ where the uncertainties in $L_{X,LMXB}$ (S) and $L_{X,HMXB}$ (S) are negligible. Given the $L_X$ bin size (0.5 dex), this error cannot change our results.

The error in $L_{X,PL}$ (S) is harder to estimate. Assuming that a single power-law (PL) represents the spectral shape of integrated emission of AGNs and X-ray binaries, the main uncertainty is the PL photon index. The photon index is typically ~1.7 in type 1 AGNs, but it may be <1 if the source is heavily obscured, as in obscured AGNs with $N_H > 10^{22}$ $cm^{-2}$. The lower the index, the smaller the $L_{X,PL}(S)$. So, the AGN contribution in the S-band is minor, even if it is significant in the H-band. When the photon index is close to 1.7 for unobscured AGNs, $L_{X,PL}(S)$ will be significant. However, we do not expect many strong unobscured AGNs in our galaxy sample because optically identified QSOs are already excluded (Kim et al. 2023a). Since we have no prior knowledge, we take a middle value of 1.4 in this work. To evaluate the impact of the photon index uncertainty, we repeated our measurements with index = 1.2 and 1.6. With the lower index (1.2), the AGN component is minor, and the overall trends described in section 5 remain the same (the left panel of Figure B-1). With the higher index (1.6), all the XLF points are shifted to the lower left as expected (the right panel of Figure B-1). However, the overall trends still remain valid.

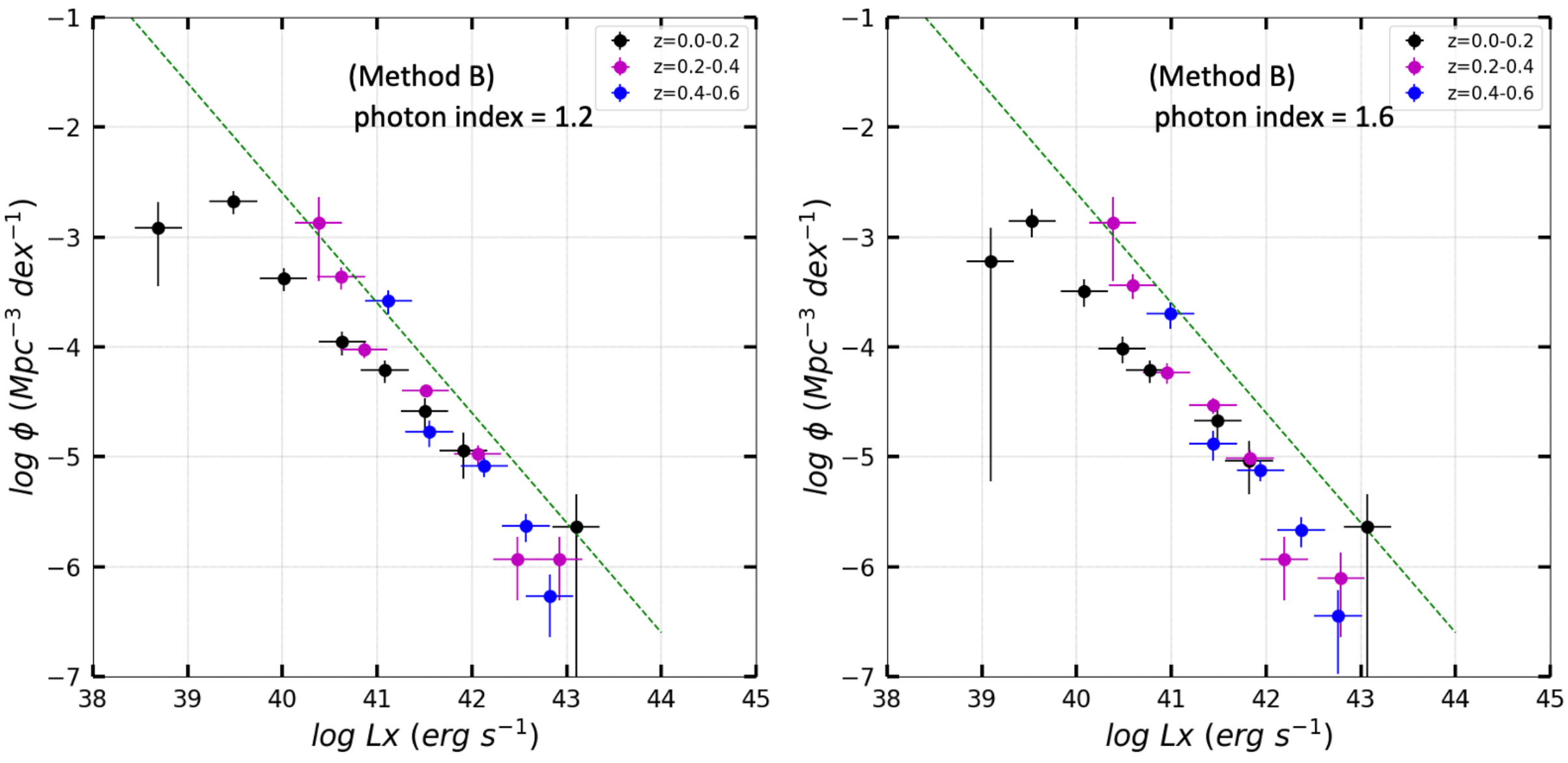

Figure B-1. same as the right panel of Figure 11, except the assumed photon-index is (left) 1.2 and (right) 1.6.

Finally, we consider the possibility that $L_{X,GAS}$ (S) is still positive but much lower after subtracting the non-gas component, moving to a different $L_X$ bin. To show how many galaxies have very low $L_{X,GAS}$ (S), we show the histograms of the ratios of $L_{X,GAS}$ (S) to Lx (S) for the two methods in Figure B-2. For Method A (left panel), 1% of galaxies have very low ratios with $L_{X,GAS}$ (S) / Lx(S) < 0.3 (or 0.5 dex = the Lx bin width). For Method B (right panel), 12% have very low ratios with $L_{X,GAS}$ (S) / Lx(S) < 0.3. Most galaxies (90% or more) remain in the same $L_X$ bin after excluding the X-ray binary or power-law component. Note that if the non-gas component is too large, $L_{X,GAS}$ (S) becomes negative and is excluded in XLF. To confirm its impact on the XLFs, we also rebuilt the XLFs after excluding them and found no significant differences.

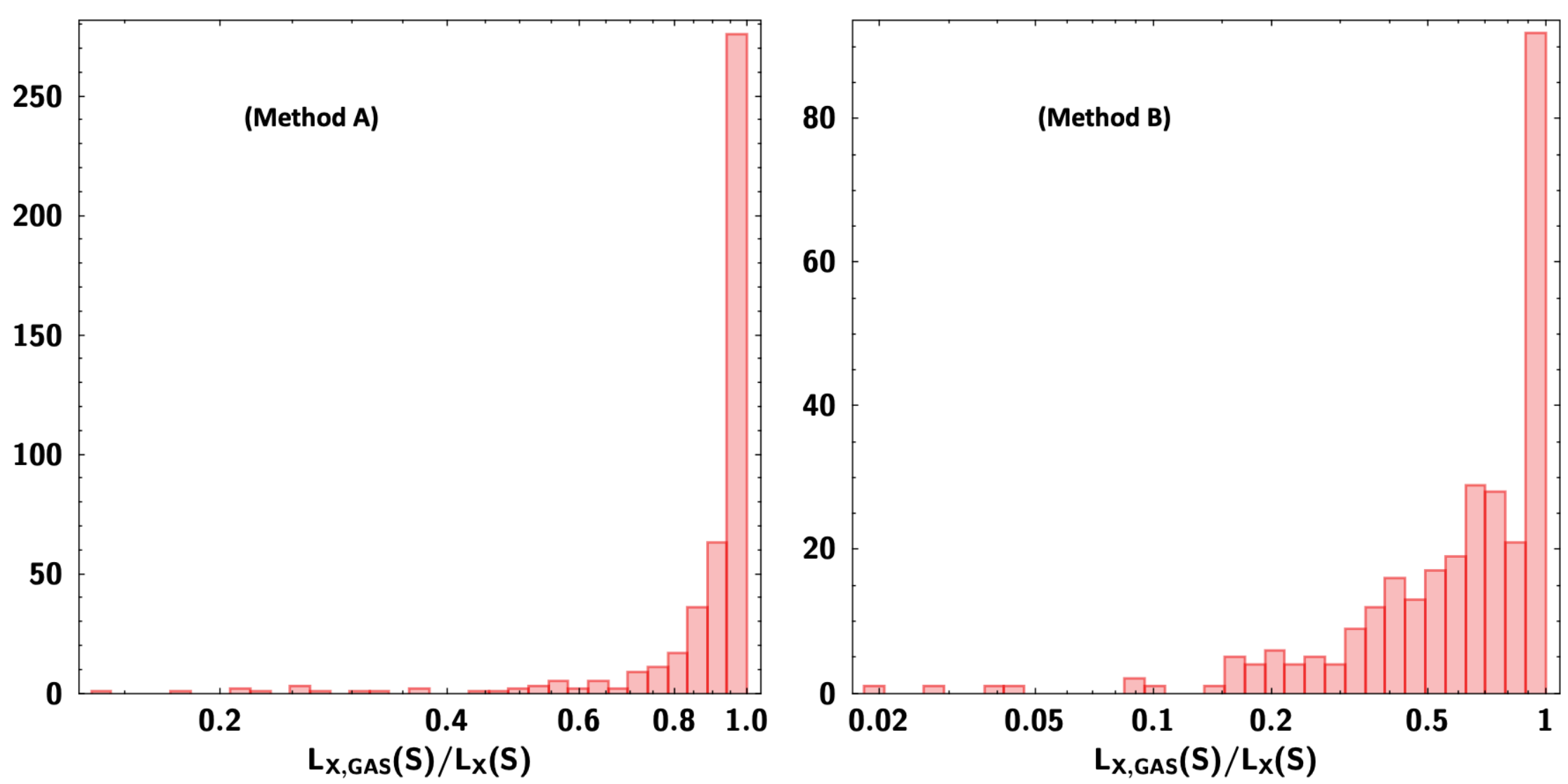

Figure B-2. Histograms of $L_{X,GAS}$ (S) to $L_X$ (S). In Method A, $L_{X,GAS}$ (S) = $L_X$ (S) – ($L_{X,LMXB}$ (S) + $L_{X,LMXB}$ (S)) and in Method B, $L_{X,GAS}$ (S) = $L_X$ (S) – $L_{X,PL}$ (S).

## Acknowledgments

We have extracted archival data from the Chandra Data Archive and the Chandra Source Catalog version 2. The data analysis was supported by the CXC CIAO software and CALDB. We have used the NASA NED and ADS facilities. This work was supported by the Chandra GO grants (AR1- 22012X) and NASA contract NAS8–03060 (CXC).